\documentclass[twocolumn]{aastex631}

\usepackage{scicite}
\usepackage{graphicx}
\usepackage[inline]{trackchanges}
\usepackage{breqn}
\usepackage{xcolor}

\newcommand{\pd}[2]{\frac{\partial #1}{\partial #2}}
\newcommand{\der}[2]{\frac{d #1}{d #2}}

\begin{document}

\title{The Limited Role of the Streaming Instability During Moon and Exomoon Formation}

\author[0000-0001-5014-0448]{Miki Nakajima}
\affiliation{Department of Earth and Environmental Sciences, University of Rochester, P.O. Box 270221, Rochester, NY 14627, USA.}
\affiliation{Department of Physics and Astronomy, University of Rochester, P.O. Box 270171, Rochester, NY 14627, USA.}

\author{Jeremy Atkins}
\affiliation{Department of Physics and Astronomy, University of Rochester, P.O. Box 270171, Rochester, NY 14627, USA.}


\author[0000-0002-3771-8054]{Jacob B. Simon}
\affiliation{Department of Physics and Astronomy, Iowa State University, Ames, IA 50010, USA.}

\author[0000-0003-1280-2054]{Alice C. Quillen}
\affiliation{Department of Physics and Astronomy, University of Rochester, P.O. Box 270171, Rochester, NY 14627, USA.}



\begin{abstract}
It is generally accepted that the Moon accreted from the disk formed by an impact between the proto-Earth and impactor, but its details are highly debated.  
Some models suggest that a Mars-sized impactor formed a silicate melt-rich (vapor-poor) disk around Earth, whereas other models suggest that a highly energetic impact produced a silicate vapor-rich disk. Such a vapor-rich disk, however, may not be suitable for the Moon formation, because moonlets, building blocks of the Moon, of 100 m-100 km may experience strong gas drag and fall onto Earth on a short timescale, failing to grow further.
This problem may be avoided if large moonlets ($\gg 100$ km) form very quickly by streaming instability, which is a process to concentrate particles enough to cause gravitational collapse and rapid formation of planetesimals or moonlets. 
Here, we investigate the effect of the streaming instability in the Moon-forming disk for the first time and find that this instability can quickly form $\sim 100$ km-sized moonlets. However, these moonlets are not large enough to avoid strong drag and they still fall onto Earth quickly.
This suggests that the vapor-rich disks may not form the large Moon, and therefore the models that produce vapor-poor disks are supported. This result is applicable to general impact-induced moon-forming disks, supporting the previous suggestion that small planets ($<1.6 R_\oplus$) are good candidates to host large moons because their impact-induced disks would be likely vapor-poor. We find a limited role of streaming instability in a satellite formation in an impact-induced disk, whereas it plays a key role during planet formation.  
\end{abstract}



\section{Introduction} 
\label{sec:intro}
\subsection{Review of the Moon-formation hypothesis}
\label{sec:review}
The origin of the Earth's Moon is a long-standing open problem in planetary science. While it is accepted that the Moon formed from a partially vaporized disk generated by a collision between the proto-Earth and an impactor approximately 4.5 billion years ago (the ``giant impact hypothesis'') \citep{HartmannDavis1975,CameronWard1976}, the details, such as the impactor radius and velocity, are actively debated. According to the canonical model, the proto-Earth was hit by a Mars-sized impactor \citep{CanupAsphaug2001}. This model can successfully explain several observations, such as the mass of the Moon, the angular momentum of the Earth-Moon system, and the small lunar core.  Moreover, this model could potentially explain the observation that the Moon is depleted volatiles, if they escaped during the impact or subsequent processes (see reviews by \citealt{Halliday2004, Canupetal2023}). However, this model fails to explain the Earth and Moon's nearly identical isotopic ratios (e.g., Si, Ti, W, O, \citealt{Armytageetal2012, Zhangetal2012, Toubouletal2015, Thiemensetal2019, Thiemensetal2021, Krujiferetal2021}). This is because the disk generated by an oblique Mars-sized impactor is primarily made of the impactor materials \citep{Canup2004, Kegerreisetal2022, Hulletal2024}, which likely had different isotopic ratios from Earth, unless the inner solar system was well mixed and homogenized \citep{Dauphas2017} or equilibrated in the disk phase \citep{PahlevanStevenson2007, Locketal2018}. 
In contrast, more energetic impact models may solve this problem by mixing the proto-Earth and impactor. These energetic models include an impact between two half Earth sized objects \citep{Canup2012}, an impact between a rapidly rotating proto-Earth and a small impactor \citep{CukStewart2012}, and general impact events that involve high energy and high angular momentum, which creates a doughnut-shaped vapor disk connected to the
Earth (a so-called Synestia) \citep{Locketal2018}. 
In these scenarios, the disk and the proto-Earth compositions could have been mixed well, potentially solving the isotopic problem.
Other proposed scenarios include (a) the Moon formation by multiple small impacts  \citep{Rufuetal2017}, (b) hit-and-run collisions that mix the proto-Earth and the Moon more than the canonical model does \citep{Asphaugetal2021}, and (c) an impact between the proto-Earth with a surface magma ocean, which leads to more contribution from Earth to the disk than the canonical scenario \citep{Hosonoetal2019}.


All of these models have potential explanations for the isotopic similarity, but each model faces challenges to explain other constraints. For example, the half-Earths, high angular momentum, and high energy models predict a much higher angular momentum of the Earth-Moon system than that of today \citep{CukStewart2012, Canup2012}. The excess of the angular momentum may or may not be removed easily \citep{CukStewart2012, Cuketal2016, Wardetal2020, Rufuetal2017, RufuCanup2020, Cuketal2021}. Another constraint may come from the Earth's interior; the deep portion of the Earth still retains solar neon \citep{YokochiMarty2004, WilliamsMukhopadhyay2019} and helium isotopic ratios \citep{Williamsetal2019}, which are distinct from the rest of the mantle. These noble gasses may have been delivered to Earth's mantle during the planet formation phase. This suggests that Earth's mantle developed chemical heterogeneity during Earth's accretion, when the protoplanetary disk was still around, and that the Earth's mantle has never been completely mixed, even by the Moon-forming impact. Previous work suggests that preservation of the mantle heterogeneity can be explained by the canonical model, but not by the energetic models, because these energetic impacts tend to mix the mantle \citep{NakajimaStevenson2015}. It should be noted, however, the possibility that the neon and helium heterogeneity developed after the Moon-formation has been discussed \citep{Bouhifdetal2020}, assuming delivery of these noble gases to the core was efficient. 

Moreover, due to recent analytical capability, small isotopic differences between Earth and the Moon have been observed (e.g., K, O, W, V, Cr, and others, \citealt{WangJacobsen2016, Wiecheretal2001, Youngetal2016, Toubouletal2015, Kruijeretal2015, Thiemensetal2019, Nielsenetal2021, Sossietal2018}). Some isotopic difference, such as the Moon's enrichment of heavy K isotopes compared to those of Earth, can be explained by isotopic fractionation during the Moon accretion process due to liquid-vapor phase separation \citep{WangJacobsen2016, NieDauphas2019, Charnozetal2021}. Observed small oxygen isotopic differences between Earth and the Moon may suggest that the Moon still records the impactor component \citep{Canoetal2020}. This suggestion may not be compatible with the energetic impact models because these impacts are so energetic that Earth and the Moon would be efficiently mixed. Whether the proposed models for the lunar origin can explain the observation that the Moon is depleted in volatiles is actively debated \citep{Canupetal2015, Dauphasetal2015, NakajimaStevenson2018, Dauphasetal2022, NieDauphas2019, Sossietal2018, Charnozetal2021, HallidayCanup2022, Canupetal2023}.


\subsection{Gas drag problem in a vapor-rich disk}
\label{sec:disk_vapor}
Another key constraint that has not been discussed until recently is the vapor mass fraction of the disk (VMF), which sensitively depends on the impact model. Less energetic impacts, such as the canonical model and the multiple impact model, produce relatively small vapor mass fractions (VMF$\sim 0.2$ for the canonical model, \citealt{NakajimaStevenson2014}), and $\sim 0.1-0.5$ for the multiple impact model, \citealt{Rufuetal2017}), while more energetic models, such as the half-Earths model and synestia model, produce nearly pure vapor disks (VMF$\sim 0.8-1$, \citealt{NakajimaStevenson2014}). The VMF of the Moon-forming disk can significantly impact the Moon accretion process; if the initial Moon-forming disk is vapor-poor (liquid-rich), moonlets can quickly form by gravitational instability from a liquid layer in the disk midplane outside the Roche radius \citep{ThompsonStevenson1988, SalmonCanup2012}. These moonlets eventually accrete to the Moon in 10s to 100s of years \citep{ThompsonStevenson1988, SalmonCanup2012, Locketal2018}.  In contrast, an initially vapor-rich disk needs to cool until liquid droplets emerge before moonlet accretion begins. Growing moonlets in the vapor-rich disk experience strong gas drag from the vapor \citep{Nakajimaetal2022}. The gas drag effect is strongest when the gas and moonlet are coupled to an adequate degree and is sensitive to the moonlet radius. It is strongest when a moonlet radius $R_p$ of a few km \citep{Nakajimaetal2022}. In contrast, much smaller moonlets are completely coupled with gas and much larger moonlets are completely decoupled from the gas, experiencing weaker gas drag effect. 
As a result, $\sim$km-sized moonlets lose their angular momentum and inspiral onto the Earth within one day \citep{Nakajimaetal2022}, a much shorter timescale than that required for lunar formation. 

This same problem was a major challenge to planet formation in the protoplanetary disk \citep{Adachietal1976, Weidenschilling1977}. Here, we quickly review the gas drag problem in the protoplanetary disk. The radial velocity of a particle in a disk is (see Equation (\ref{eq:vr}) for the derivation) \citep{Armitage2010, TakeuchiLin2002}, 
\begin{equation}
v_r = \frac{\tau^{-1}_{\rm f}v_{r, {\rm g}}-2\eta v_{\rm K}}{\tau_{\rm f} +\tau_{\rm f}^{-1} },
\label{eq:vr_main}
\end{equation}
where $\tau_{\rm f}=\Omega t_{\rm f}$ is the dimensionless stopping time \citep{Armitage2010}, $\Omega$ is the Keplerian angular velocity, $t_{\rm f}$ is the friction time (see further descriptions below), $v_{r, {\rm g}}$ is the gas velocity, $\eta$ is the pressure gradient parameter described as
\begin{equation}
\eta = -\frac{1}{2} \left(\frac{c_{\rm s}}{v_{\rm K}}\right)^2 \frac{ \partial \ln p_{\rm g}}{\partial \ln r}, 
\label{eq:eta}
\end{equation}
where $c_{\rm s}$ is the sound speed, and $v_{\rm K}$ is the Keplerian velocity,  
 $p_{\rm g}$ is the gas pressure, and $r$ is the radial distance. $v_r$ is largest when $\tau_f = 1$. The friction time is the time until the particle and gas reaches the same velocity and is defined as $t_{\rm f} = m{v_{\rm rel}}/F_{\rm D}$, where $m$ is the particle mass, $v_{\rm rel}$ is the relative azimuthal velocity between the gas and particle, and $F_{\rm D}$ is the drag force. In the protoplanetary disk, the gas drag for a particle is generally written as $F_{\rm D}=-\frac{C_{\rm D}}{2}{\pi \rho_{\rm g,0} R_{\rm p}^2 v_{\rm rel}^2}$, where $C_{\rm D}$ is the gas drag coefficient and $\rho_{\rm g,0}$ is the initial gas density at the midplane, and $R_{\rm p}$ is the particle radius. The gas drag coefficients are roughly \citep{Armitage2010}
 \begin{equation}
 \label{eq:CD}
  C_D=\begin{cases}
    24 {\rm Re}^{-1}, & \text{${\rm Re}<1$ (Stokes regime)} \\
    24{\rm Re}^{-0.6}, & \text{$1<{\rm Re}<800$ (Transition regime)}\\
    0.44, & \text{${\rm Re}>800$ (Newton regime)}\\
  \end{cases}
\end{equation}
where Re is the Reynolds number. Assuming that $v_{\rm rel}\sim \eta v_{\rm K}$ (see Equation \ref{eq:v_phi}), $\nu \sim c_s \lambda$, where $\nu$ is the kinematic viscosity, $\lambda$ is the mean free path ($\lambda=\frac{k_{\rm B} T}{\sqrt{2}\pi d^2 p_{\rm g,0}}$, where $k_{\rm B}$ is the Boltzmann constant, $T$ is the temperature, $d$ is the molecular diameter, $p_{\rm g,0}$ is the gas pressure at the mid-plane), Re=$\frac{v_{\rm rel}R_p}{\nu}\sim4.26$, which is in the transition regime, in the protoplanetary disk at 1 AU, assuming $T=280$ K, $\eta=0.002$, $d = 289$ pm, $R_{\rm p}=0.52$ m (see discussion below),  $p_{\rm g,0}=\frac{\rho_{g,0} R T}{m_{\rm m}}$, and $m_{\rm m}=0.002$ kg/mol, where $m_{\rm m}$ is the mean molecular weight and $R$ is the gas constant. Here we also assume the following vertical distribution of the gas density $\rho_{\rm g}$, 
\begin{equation}
    \rho_{\rm g} = \rho_{\rm g,0} \exp\left( -\frac{z^2}{2H^2}\right),
    \label{eq:gas}
\end{equation}
where $z$ is the vertical coordinate and $H$ is the gas scale height. This leads to $\rho_{\rm g,0}=\frac{\Sigma_{\rm g}}{\sqrt{2\pi}H}$. We assume $\Sigma_{\rm g} =17000$ kgm$^{-2}$ and use the relationships of $c_s = \sqrt{RT/m_{\rm m}}$, $H=c_s/\Omega$, $\frac{\partial \ln p_{\rm g}}{\partial \ln r}=(-3 +\beta/2)$, $\beta=\frac{1}{2}$. The orbital angular frequency is $\Omega = \sqrt{GM/r^3}$, where
$M$ is the stellar mass, and $r$ is the distance from the star (for the Moon-forming disk, $M$ is the Earth mass and $r$ is the distance from Earth). These parameters are taken from previous work \citep{Carreraetal2015} (other values of $\beta$ have been discussed, such as  $\beta=\frac{3}{7}$, \citealt{ChiangYoudin2010}).

 The dimensionless stopping time becomes 
\begin{equation} 
\label{tau_f}
\tau_{\rm f} = \Omega t_{\rm f} = \Omega \frac{8}{3 C_{\rm D}} \frac{\rho_{\rm p} R_{\rm p}}{\rho_{\rm g,0} v_{\text{rel}}}=\frac{8}{3 C_{\rm D}\eta} \frac{\rho_{\rm p} R_{\rm p}}{\rho_{\rm g,0} r}.
\end{equation}
$\tau_{\rm f}=1$  when $R_{\rm p}=0.52$ m. The particle density $\rho_p=3000$ kgm$^{-3}$ is assumed. The residence time of the particle at 1 AU is 1 AU$/v_r=75$ years, where $v_r$ is the radial fall velocity of the particle (see \ref{eq1}. For simplicity, the radial gas velocity $v_{r,{\rm g}}=0$ is assumed). Thus, an approximately meter-sized particle falls toward the Sun at 1 AU within 80 years, a timescale much shorter than the planet formation timescale (several-tens of Myr). This is the so-called ``meter barrier'' problem, which was a major issue to explain planetary growth in a classical planet formation model \citep{Adachietal1976, Weidenschilling1977}.

In contrast, for the Moon-forming disk, $\tau_{\rm f} =1$  (${\rm Re}>10^{10}$) is achieved when $R_p=1.8$ km, assuming $\rho_{\rm g,0} = 40$ kg/m$^3$, $\rho_{\rm p} = 3000$ kg/m$^3$, $r=3R_\oplus$, $\eta= 0.04$, $T=5200$ K (see Section \ref{sec:modelparameters} for justifications for these parameters), and $d=300$ pm. The residence time, $r/v_r$, is approximately 1 day (1.21 day), which is much shorter than the Moon-formation timescale of several 10s of years to 100 years. This formation timescale is ultimately determined by the radiative cooling timescale, but it is model dependent. Here we provide a very simple estimate; the time scale for radiative cooling is $\frac{M_{\rm disk}(L+C_p \Delta T)}{4 \pi r^2 \sigma T_{\rm ph}^4}\sim10$ years (this is also consistent with numerical work, \citealt{Locketal2018}), where $M_{\rm disk}$ is the disk mass, $L$ is the latent heat, $C_p$ is the specific heat, $\Delta T$ is the temperature change over time, $\sigma$ is the Stefan–Boltzmann constant, and $T_{\rm ph}$ is the photosphere temperature. Here, $M_{\rm disk}\sim0.015 M_\oplus$, $L=1.2\times 10^7$ J/kg \citep{Melosh2007}, $C_p=10^3$ J/K/kg, $\Delta T=2000$ K, $T_{\rm ph}=2000$ K \citep{ThompsonStevenson1988} are assumed. However, the actual Moon-formation timescale can be longer than this for several reasons, including additional heating due to viscous spreading \citep{ThompsonStevenson1988, CharnozMichaut2015}, and radial material transport efficiency (e.g. \citealt{SalmonCanup2012}). Short Moon-formation timescale has been proposed (e.g. \citealt{MullenGammie2020, Kegerreisetal2022}), but the typical Moon-formation timescale has been estimated to range in $10-10^2$ years.

The vertical settling velocity of condensing particles is $v_{\rm settle} = \sqrt{\frac{8}{3  C_{\rm D}}\frac{\rho_{\rm p} R_{\rm p} \Omega^2 z}{\rho_{\rm g}}}$ \citep{Armitage2010}. Assuming $z\sim H$, the settling time for a particle with a radius of 2 km is 0.22 days. Determining the collision history of moonlets require conducting orbital dynamics simulations, but here we provide a rough estimate. A rough estimate of the mass doubling time of the largest moonlet in the Moon-forming disk is typically $\sim 1$ day \citep{SalmonCanup2012}. If a 2 km-sized moonlet mass doubles in a day, the radius change is very small (2km$\times 2^{1/3}=2.5$ km) and the gas drag effect remains strong on such a moonlet and this change does not prevent the moonlet from falling into Earth.
However, the actual collision time can vary depending on the local concentration of particles, which needs a detailed future study.

This gas drag effect is a problem with forming the Moon from an initially vapor-rich disk.
The Moon can still form after most of the vapor condenses, but by that time a significant portion of the disk mass could be lost \citep{Idaetal2020, Nakajimaetal2022}, which would fail to form a large moon from an initially vapor-rich disk (here we use a ``large'' moon when its mass is $\sim1$ wt\% or larger of the host planet). If this is the case, an initially vapor-rich disk may not be capable of forming a large Moon \citep{Nakajimaetal2022}. In contrast, the gas drag effect is weak for vapor-poor disks, which are generated by less energetic models, such as the canonical and multiple impact models.


\subsection{Streaming instability in the general impact-induced moon-forming disk}
\label{sec:SI_moon}
A potential solution to this vapor drag problem is forming a large moonlet very quickly (much larger than 2 km), so that the moonlet would not  experience strong gas drag. This is the accepted solution for the gas drag problem in the protoplanetary disk. The proposed mechanism is the streaming instability \citep{YoudinGoodman2005, Johansenetal2007}, where particles spontaneously concentrate in the disk, gravitationally collapsing and forming a large clump ($\sim 100$ km in size, \citealt{Johansenetal2015}). If this mechanism works for the Moon-forming disk, an initially vapor-rich disk may be able to form the Moon despite the gas drag issue. If this mechanism turns out not to work for the Moon-forming disk, it is an interesting finding as well, given that a Moon-forming disk is often treated as a miniature analogue of the protoplanetary disk. Understanding what makes these two disks differ would deepen our understanding of planet and satellite formation processes.

Moreover, knowing whether the streaming instability can affect moon formation processes informs our understanding of moon formation in the solar and extra-solar systems. Moon formation in an impact-induced disk is common in the solar system (e.g. Martian moons, \citealt{Craddock2011}, Uranian moons, \citealt{Slatteryetal1992}, and Pluto-Charon \citealt{Canup2005}). While there are no confirmed exomoons (moons around exoplanets) to date \citep[e.g.][]{CasseseKipping2022}, impact-induced exomoons should be common because impacts are a common process during planet formation \citep{Nakajimaetal2022} and because impacts in extrasolar system may have been already observed (e.g. \citealt{Mengetal2014, Bonomoetal2019, Thompsonetal2019, Kenworthyetal2023}). If streaming instability operates in these disks, it can affect what types of planets can host exomoons, which can be compared with future exomoon observations. 

It should be also noted that other instability, such as secular gravitational instability (GI) (e.g. \citealt{Youdin2011, Takahashietal2014,  Tominagaetal2018}) and two-component viscous GI (TVGI) \citep{Tominagaetal2019}, have been discussed as a mechanism for planetesimal and dust ring formation. The GI occurs when dust-gas interaction reduces the rotational support of the rotating disk, which leads to dust concentration. The TVGI is an instability cased by dust–gas friction and turbulent gas viscosity. These instabilities can lead to clump formation even if the disk is self-gravitationally stable. Implications of these instabilities are beyond the scope of this paper.



\subsection{Motivation of this work}
\label{sec:motivation}
The goal of this work is to investigate whether the streaming instability can form moonlets that are large enough to avoid strong gas drag from the vapor-rich disk. The result would constrain the Moon-formation model as well as general impact-induced models in the solar and extrasolar systems. 
We conduct hydrodynamic simulations using the code Athena \citep{Stoneetal2008, BaiStone2010a}. We first conduct 2D simulations to identify the section of parameter space that leads to streaming instability (SI). Subsequently, we conduct 3D simulations with self-gravity to identify the size distribution of moonlets. 
Lastly, we identify the lifetime of the moonlets formed by streaming instability to investigate whether it is possible to form a large moon from an initially vapor-rich disk. For the general impact-induced disks, we consider ``rocky'' and ``icy'' disks, where these disks form by collisions between rocky planets (with silicate mantles and iron cores) and between icy planets (with water ice mantles and silicate cores), respectively.


\section{Method} 
\subsection{Athena}
We use the Athena hydrodynamics code, which solves the equations of hydrodynamics using a second-order accurate Godunov flux-conservative approach \citep{Stoneetal2008}. We use the configuration of Athena that couples the dimensionally-unsplit corner transport upwind method \citep{Colella1990} to the third-order in space piecewise parabolic method by Colella \& Woodward \citep{ColellaWoodward1984} and calculates the numerical fluxes using the HLLC Riemann solver \citep{Toro1999}. We also integrate the equations of motion for particles following Bai \& Stone \citep{BaiStone2010a} and include particle self-gravity for 3D simulations following the particle-mesh approach described in previous work \citep{Simonetal2016}. Orbital advection is taken into account following previous work \citep{BaiStone2010a, BaiStone2010b}.
Our setup is the local shearing box approximation in which a small patch of the disk is corotating with the disk at the Keplerian velocity \citep{StoneGardiner2010}. The local Cartesian frame is defined as ($x, z$) for 2D and ($x, y, z$) for 3D simulations, with $x$ as the radial coordinate from the planet and $z$ is parallel to the planetary spin axis, and $y$ is in the direction of orbital rotation. 

In the Athena code, we solve the following equations for our simulations  \citep{BaiStone2010a, Simonetal2016, Lietal2019}: 
\begin{equation}
    \pd{\rho_{\rm g}}{t} + \nabla \cdot (\rho_{\rm g} \mathbf{u}) = 0,
    \label{eq:mass}
\end{equation}

\begin{equation}
    \pd{\rho_{\rm g} \mathbf{u}}{t} + \nabla \cdot (\rho_{\rm g} \mathbf{uu} + p_g\mathbf{I}) = 3 \rho_{\rm g} \Omega^2 \mathbf{x}  - \rho_{\rm g} \Omega^2  \mathbf{z} - 2 \rho_{\rm g} \mathbf{\Omega} \times  \mathbf{u}  
    + \rho_{\rm p} \frac{\overline{\mathbf{v}} - \mathbf{u}}{t_{\text{\rm f}}},
\label{eq:momentum}
\end{equation}

\begin{equation}
    \der{\mathbf{v}_i}{t}
    =   -2 \eta v_{\rm K} \Omega \hat{\mathbf{x}} + 3 \Omega^2 x_i  \hat{\mathbf{x}}\\
     - \Omega^2 z_i \hat{\mathbf{z}}
     - 2  \mathbf{\Omega} \times  \mathbf{v_i}  - \frac{\mathbf{v}_i- \mathbf{\overline{u}}}{t_{\rm f}} +  \mathbf{a_{\rm g}}.
\label{eq:particle}
\end{equation}
The first two equations specify mass and momentum conservation for the gas, respectively, while the third equation represents the motion of a particle $i$ coupled with the gas. Here, $\mathbf{u}$ is the velocity of the gas and $\mathbf{I}$ is the identity matrix. $\mathbf{\overline{v}}$ is the mass-weighted averaged particle velocity in the fluid element, assuming that particles can be treated as fluid \citep{BaiStone2010a}. 
The terms of the right hand side of Equation (\ref{eq:momentum}) are radial tidal forces (gravity and centrifugal force), vertical gravity, and the Coriolis force, and the feedback from the particle to the gas. 
In Equation (\ref{eq:particle}), the first term on the right hand side describes a constant radial force due to gas drag. $\mathbf{v}_i$ is the particle velocity, 
$\hat{\mathbf{x}}$ and $\hat{\mathbf{z}}$ represent the unit vectors in the $x$ and $z$ directions, $x_i$ and $z_i$ are the values of $x$ and $z$ for the particle $i$. $\mathbf{\overline{u}}$ is the gas velocity interpolated from the grid cell centers to the location of the particle. The second, third, and forth terms are radial and vertical tidal forces and the Coriolis force.
 $\mathbf{a_g}$ is the acceleration due to self-gravity, which is considered only in 3D simulations. $\mathbf{a_g} = - \nabla \Phi_p$, where $\Phi_p$ is the gravitational potential and satisfies Poisson's equation, $\nabla^2 \Phi_p = 4 \pi G \rho_p$.
To reduce computational time, the particles are organized into ``superparticles'', each representing a cluster of individual particles of the same size. 
In the code units, we normalize $\Omega = c_s = H = \rho_{g,0} = 1$.
%
The gas and particle initial distributions are described as Equation (\ref{eq:gas}) and 
\begin{equation}
    \rho_{\rm p} = \frac{\Sigma_{\rm p}}{\sqrt{2\pi}H_{\rm p}} \exp\left( -\frac{z^2}{2H_{\rm p}^2}\right),
    \label{eq:particle_z}
\end{equation}
respectively, where $H_{\rm p}$ is the scale height of particles and is set to 0.02$H$ \citep{BaiStone2010a} and $\Sigma_{\rm p}$ is the particle surface density.
The system uses an isothermal equation of state $P = \rho_g c_s^2$, and the particles are distributed uniformly in the $x$ and $y$ direction and normally in the $z$ direction (Equation \ref{eq:particle_z}). 
The computational domains are $0.2H \times 0.2H$ in 2D simulations and $0.2H \times 0.2H \times 0.2H$ in 3D simulations, with all-periodic boundary conditions. The resolution for 2D is 512 $\times$ 512 and each grid cell has 1 particle ($512\times512=262,144$ particles) .
 The resolution for 3D is $10\times 128^3$ ($=21$ million particles in total). Previous work shows that the resolution of $128^3$ can produce the large clump size distribution well compared to those with $256^3$ and $512^3$ \citep{Simonetal2016} and therefore this 3D resolution is sufficient to resolve large clumps, which are the focus of this work. The initial particle size is assumed to be constant. Variable initial particle sizes could affect the growth speed \citep{Krappetal2019} and the concentrations of particles depending on their sizes \citep{Yangetal2021}, but it is not known to affect the largest clump size formed by streaming instability.

\subsection{Clump detection in 2D and 3D}
\label{sect:clump}
After we conduct 2D simulations, we identify filaments forming in the disk in order to constrain the parameter space favorable for the streaming instability. We use the \emph{Kolmogorov-Smirnov} (KS) method, which is based on previous work \citep{Carreraetal2015}. For each time step in a given time window ($25/\Omega$ in our case), we compute the particle surface density $\Sigma_p(x)=\int_{-0.1}^{0.1} \rho_p(x,z)\,dz$ and average it over the time window to give $\left< \Sigma_p \right>_t$. We then sort the values in $\left< \Sigma_p \right>_t$ from highest to lowest, and compute the cumulative distribution of this sorted dataset. Finally, we use the Kolmogorov-Smirnov test to output a $p$-value that measures the likelihood that the underlying cumulative distribution was linear:
\begin{align}
    Q(z) &= 2 \sum_{j=1}^\infty (-1)^{j-1} e^{-2 z^2 j^2}\\
    p &= Q(D \sqrt{n})
\label{Pvalue}
\end{align}
where $D$ is the maximum distance between the data and linear cumulative distributions, and $n$ is the number of data points. The higher the $p$ is, the more homogeneous the system is and therefore filament formation is unlikely, whereas small $p$ indicates filament formation is more likely. Here we assume that filament formation is very likely when $p<0.10$, the same as in previous work \citep{Carreraetal2015}.

For self-gravitating clump detection in our 3D simulations, we use PLanetesimal ANalyzer (PLAN) \citep{Lietal2019}. This is a tool to identify self gravitating clumps specifically made for Athena output. The density of each particle is assessed based on nearest particles. Particles with densities higher than a threshold are associated with neighbouring dense particles until a density peak is achieved. In contrast, particle groups with a saddle point less than a threshold remain separated. Detailed descriptions are found in previous work \citep{Lietal2019}.

\subsection{Model Parameters}
\label{sec:modelparameters}

The main parameters for the Athena simulations are (1) the dimensionless stopping time $\tau_{\rm f}$ (the Newton regime, see Section \ref{sec:disk_vapor} and \citealt{Nakajimaetal2022}),
(2) the normalized pressure parameter $\Delta = \eta v_{\rm K}/c_{\rm s}$, (3) the ratio of the particle surface density to the gas surface density $Z = \Sigma_{\rm p} / \Sigma_{\rm g}$ (VMF=$\frac{1}{1+Z}$), and (4) the normalized gravity $\tilde{G} \equiv  4 \pi G \rho_{\rm g,0}/\Omega^2$ for 3D simulations, where $G$ is the gravitational constant.
A small value of $\tau_{\rm f}$ indicates a small particle radius $R_{\rm p}$, where the particle is well coupled with the gas, whereas a large value of $\tau_{\rm f}$ corresponds to a large value of $R_{\rm p}$, which is more decoupled from the gas. A large value of $\Delta$ corresponds to a large pressure gradient and quicker radial infall. $\tilde{G}$ represents the strength of self-gravity.  
The parameter space we are exploring for our 2D simulations is $\tau_f = 10^{-3}, 10^{-2}, 10^{-1}, 10^0, 10^1, \Delta = 0.1, 0.2, 0.3, 0.4, 0.5$ and $Z = 0.05, 0.1$. For 3D simulations, we use $Z=0.1, 0.3$ and $\tilde G = 0.1788$ and $0.5898$, where the former $\tilde{G}$ value corresponds to slightly cooler temperature ($4700$ K) while the latter corresponds to hotter temperature ($5200$ K). 
Here, we justify the choice of these parameters.
This range of $\tau_f$ corresponds to the particle radius of 2 m  and 20 km (see Equation \ref{tau_f}).
 The global disk structures have been calculated based on hydrodynamic simulations in previous work. The overall disk mass of the Moon-forming disk is typically a few percent of Earth, depending on the impact model ($M_D/M_L=$1.35-2.80, where $M_D$ is the disk mass and $M_L$ is the lunar mass, \citealt{NakajimaStevenson2014}). The mid-plane disk temperature ranges from 3000 K to 7000 K and the radial range of the disk  is $\sim$1-8 $R_\oplus$ (see Figure 5 in \citealt{NakajimaStevenson2014}). The disk temperature is $\sim4000-5500$ K at $r=3 R_\oplus$.  For the general rocky and icy impact-induced disks, the disk temperature can vary, but typically in the range of thousands of K for vapor-rich disks \citep{Nakajimaetal2022}. The pressure gradient can vary, but the typical value of $\eta$ is $\sim 0.02-0.06$ based on impact simulations \citep{Nakajimaetal2022}. In the Moon-forming disk, the value of $Z$ increases as the disk cools (on the timescale of $10-10^2$ years). In other words, $Z$ can be zero initially in an energetic Moon-forming impact model (e.g. \citealt{Locketal2018}) and eventually becomes infinity as the disk materials condense and the gas disappears. Thus, picking a value of $Z$ means that we are seeing physics at a specific time.
 Since we are primarily interested in high-vapor disks (i.e. an early phase of the disk), we explore the mass ratio range $Z \in [0.05, 0.1]$ for our 2D simulations and $Z \in [0.1, 0.3]$ for our 3D simulations. We use the large value of $Z=0.3$ as a sensitivity test. Higher values of $Z$ can be achieved as the disk cools, but we focus on small values of $Z(\leq  0.3)$ for several reasons. First, at a larger value of $Z(>0.3)$, the conventional gravitational instability in the liquid part of the disk can occur to form moonlets. At $Z=0.3$, VMF=$\frac{1}{Z+1}=0.76$ and the total ($\Sigma_{\rm p} + \Sigma_{\rm g}$) surface density at $r\sim 3R_\oplus$ is $\sim 10^8$ kg m$^{-2}$ in energetic models, which means that $\Sigma_{\rm p} = 0.76 \times 10^8 $kg m$^{-2}=7.6\times10^7$ kg m$^{-2}$. Previous work suggests that when the liquid (melt) layer's thickness reaches 5-10 km (or equivalent of a few $10^7$ kg m$^{-2}$), gravitational instability can happen in the melt layer \citep{MachidaAbe2004}. Whether streaming instability occurs at the same time, or whether streaming instability affects the gravitational instability in the Moon-forming disk are unknown and have not been explored in previous studies. Under these circumstances, the important of streaming instability becomes unclear. Additionally, these streaming instability simulations have been conducted at low $Z$ values $(\leq 0.1)$ in previous work to reproduce conditions in the protoplanetary disks (e.g. \citealt{Abodetal2019, LiYoudin2021}) and simulations with high $Z(>0.3)$ values have not been fully tested. For these reasons, we focus on relatively small values of $Z$ in this study.

For the set of Athena simulations, we focus on reproducing two sets of the Moon-forming disk thermal profile; assuming $M=M_\oplus$ and $r=3R_\oplus$, where $M_\oplus$ is the Earth mass and $R_\oplus$ is the Earth radius, the gas pressure at the midplane around $3R_\oplus$ is $\approx 12$ MPa, $T=4700$ K, and $m_{\rm m}=30$ g/mol. This makes the density $\rho_{\rm g,0}=12.13$ kg m$^{-3}$ for an ideal gas, $c_{\rm s}=1140$ m/s, $v_{\rm K}=4562$ m/s, $\Omega=2.38\times 10^{-4}$ s$^{-1}$, and the scale height $H = c_s/\Omega=4.78\times10^6$ m, and $\tilde G = 0.1788$. For the higher-temperature scenario, $T=5200$ K, $\rho_{\rm g,0}=40.01$ kg m$^{-3}$, $c_{\rm s}=1200$ m/s, $H =5.03\times10^6$, and  $\tilde G = 0.5898$. These temperatures are motivated by previous hydrodynamic calculations of the Moon-forming disk formation \citep{NakajimaStevenson2014} and other parameters are calculated based on an equation of state of dunite assuming that the vapor and liquid phases are in equilibrium (MANEOS, \citealt{ThompsonLauson1972, Melosh2007}). Since we are assuming the disk is in the liquid-vapor equilibrium, higher disk temperature leads to higher vapor density at the mid-plane. For example, the gas density at the liquid-vapor equilibrium is 150 kg/m$^3$ at 6000 K and 1.8 kg/m$^3$ at 4000K for dunite, according to MANEOS (\citealt{ThompsonLauson1972, Melosh2007}).
Assuming $\eta \sim 0.02-0.06$ for the Moon-forming disk, $\Delta \sim 0.1-0.2$. However, higher values are possible \citep{Nakajimaetal2022}, and therefore we explore the range of $\Delta \sim 0.1-0.5$.

 The parameter values of $\Delta$ and $\tilde G$ are significantly different from values in the protoplanetary disk, where the typical values used are $\Delta \sim 10^{-3}-10^{-2}$ and $\tilde G \sim 0.05$ in the protoplanetary disk \citep{Carreraetal2015, Simonetal2016}. We use a fixed value of $Z$ in hydrodynamic calculations because the condensation timescale ($\sim$years) is longer than the simulation timescale (we run 2D simulations for  $\sim 100$ orbits, which correspond to 123 hours. A steady state is reached by this time). We also assume that the disk does not evolve on this short timescale.


\section{Results}
\subsection{2D Athena simulations}
\label{sec:results2D}

\begin{figure*}
  \begin{center}
\includegraphics[scale=0.075]{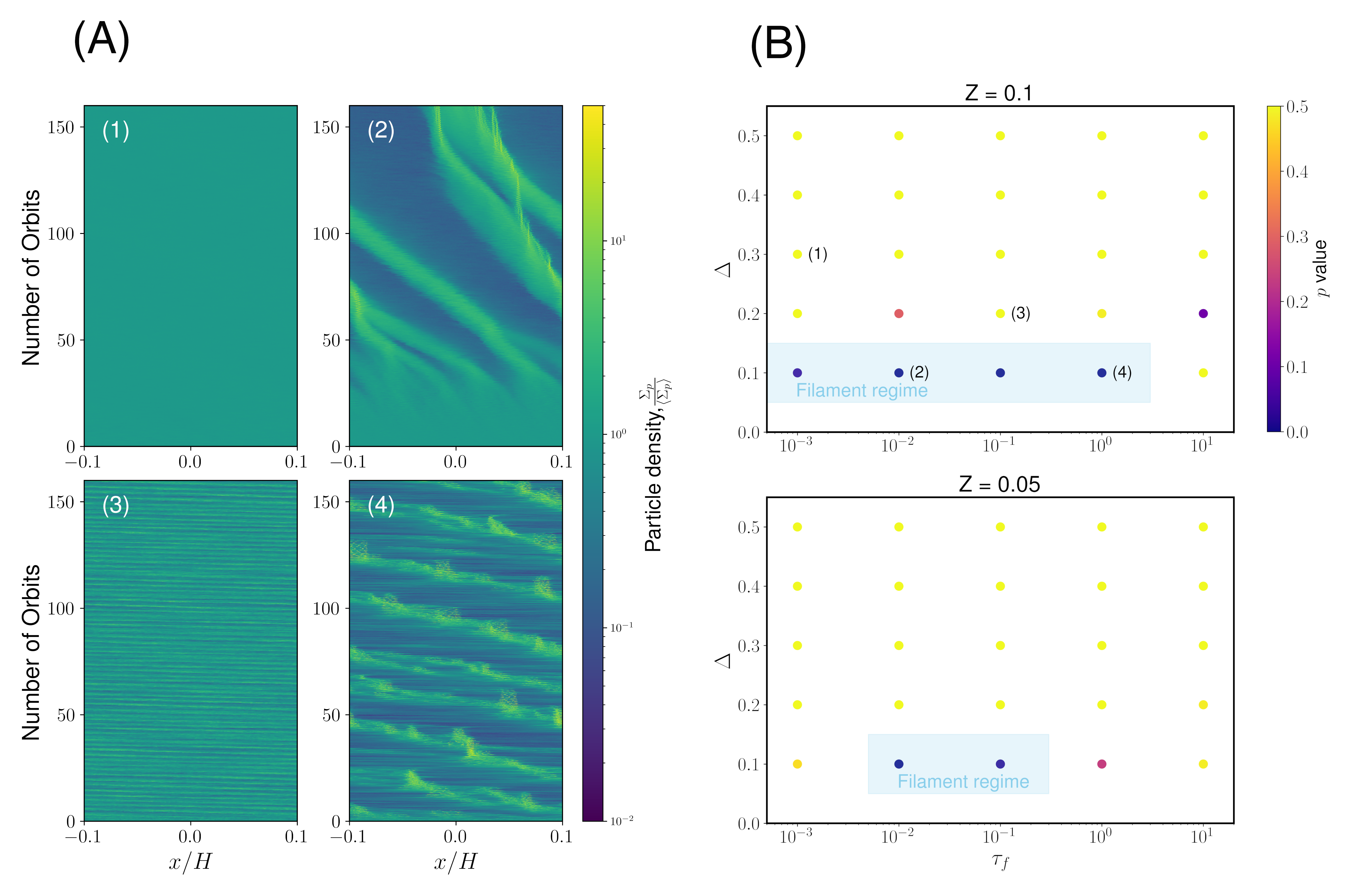}
  \end{center}
  \caption{(A) Spacetime diagram for the four cases. The input parameters are (1) $(\tau_f, \Delta)= 10^{-3}, 0.3$, (2) $ 10^{-2}, 0.1$, (3) $ 10^{-1}, 0.2$, and (4) $1, 0.1$, all at $Z=0.1$. The horizontal axis is $x$ normalized by the scale height $H$. The vertical indicates the number of orbits. The color shows $\Sigma_p/\left<\Sigma_p\right>$, where $\Sigma_p$ is the particle surface density and $\left<\Sigma_p\right>$ is the average along the $x$-axis. Filament formation occurs in cases (2) and (4), while no filament formation occurs in cases (1) and (3). (B) Result summary for $Z=0.1$ and $Z=0.05$. The colors show the $p-$value and the clumping regime ($p<0.1$) is shown in the skyblue shade. Parameters for cases 1-4 are indicated. This shows that clumping occurs only at small $\Delta$ ($\Delta = 0.1$). }
\label{fig:spacetime}
\end{figure*}

\begin{figure*}
  \begin{center}
\includegraphics[scale=0.6]{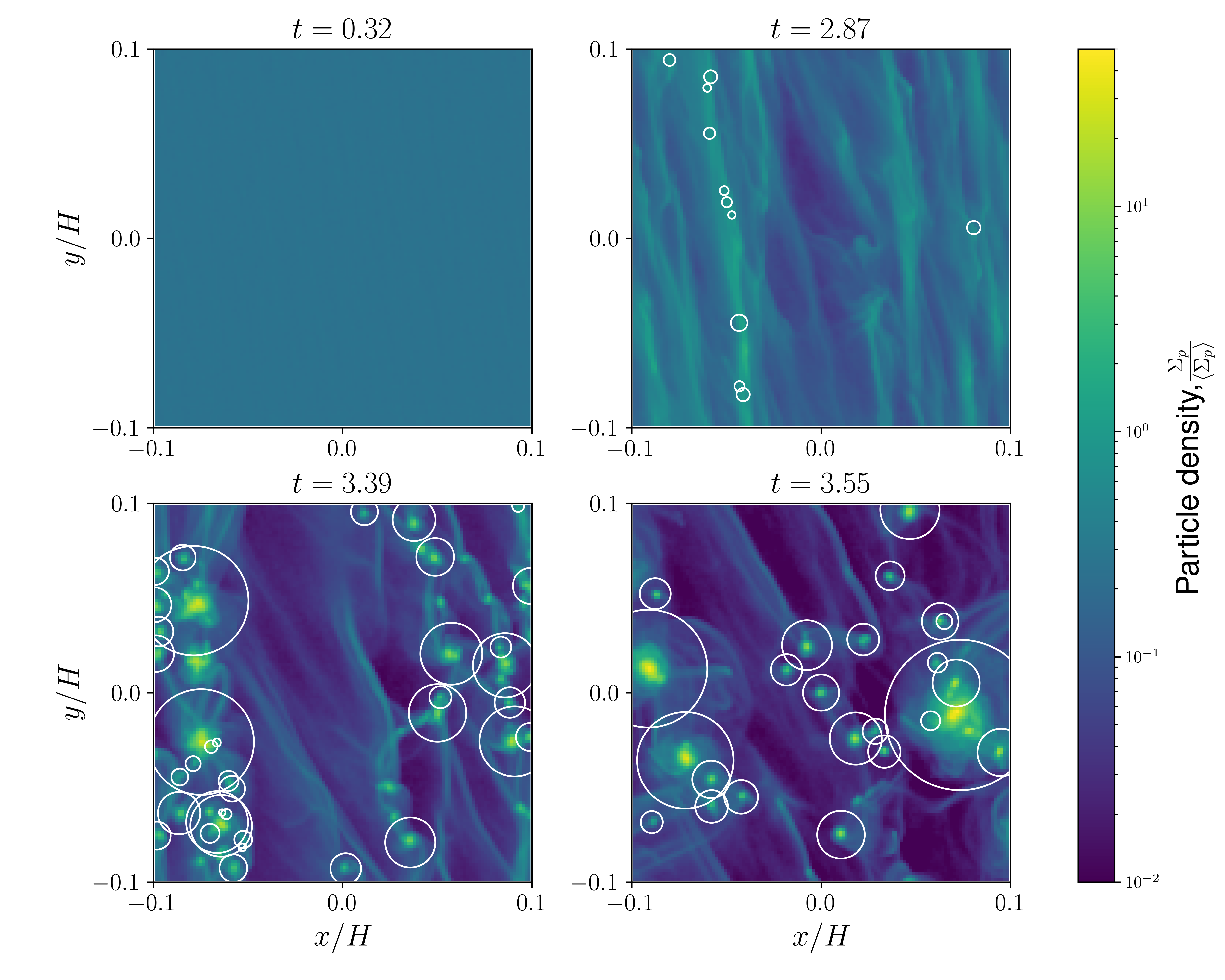}
  \end{center}
  \caption{Snapshots of a 3D simulation ($\tau_f=1, \Delta=0.1, Z=0.1$, $\tilde G=0.5898$). The horizontal and vertical axes are $x$ and $y$, normalized by the scale height $H$. The color indicates the particle density, normalized by the average density. $t$ indicates the number of orbits. The circles represent the Hill radius of each moonlet formed by streaming instability. At first ($t=0.32$), no concentration of particles is observed, but streaming instability clearly develops by $t=2.87$. After self-gravity is turned on at $t=3.18$, moonlets form by gravitational instability ($t=3.39, 3.55)$. The self-gravitating clumps are identified using PLAN (see the main text). }
\label{fig:3D}
\end{figure*}

\begin{figure*}
  \begin{center}
\includegraphics[scale=0.05]{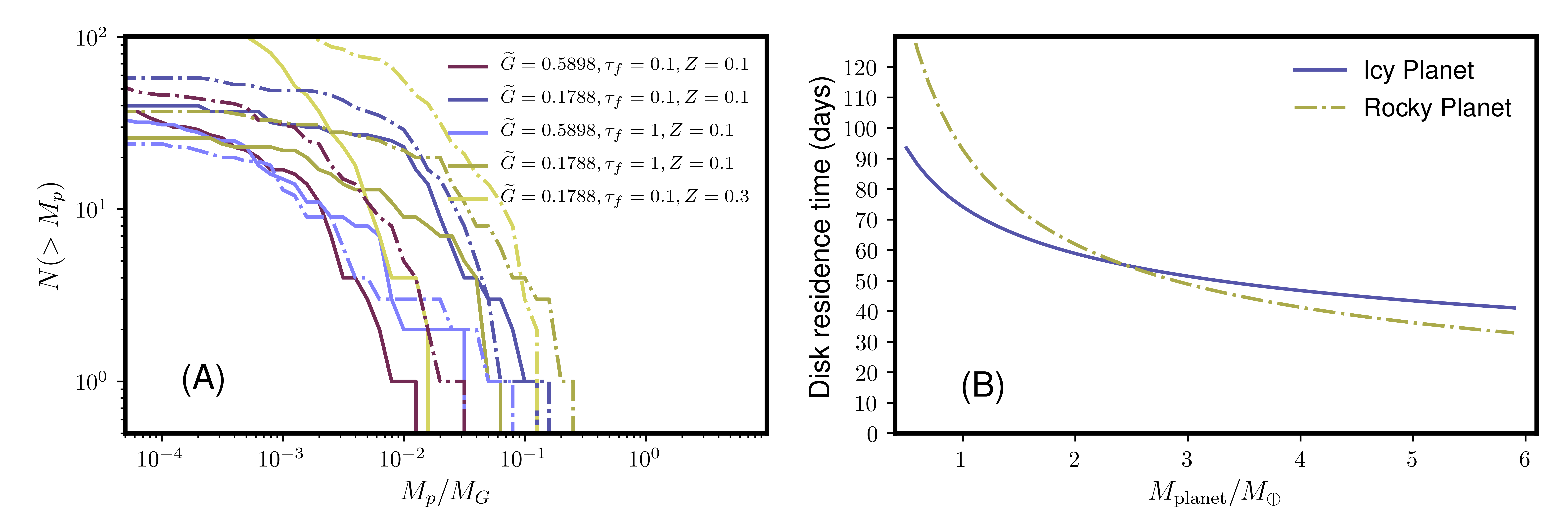}
  \end{center}
    \caption{(A) Cumulative mass distribution of moonlets formed by streaming instability at $\Delta=0.1, Z=0.1, 0.3$. The purple, dark blue, light blue, dark yellow, and light yellow lines indicate parameter values of $(\tilde{G},\tau_{\rm f}, Z)=(0.5898, 0.1, 0.1), (0.1788, 0.1, 0.1), (0.5898, 1, 0.1), (0.1788, 1, 0.1), (0.1788, 0.1, 0.3)$. The solid and dashed lines represent the mass distribution at different times (see Section \ref{sec:results3D}). The horizontal axis indicates the moonlet mass ($M_p$) normalized by $M_G$ and the vertical axis indicates the number of moonlets whose masses are larger than the given moonlet mass. (B) Residence time of a moonlet whose mass is $M_G$ formed by streaming instability  as a function of the planetary mass $M_{\rm planet}$ normalized by the Earth mass at the Roche radii. The blue solid and yellow dash-dot lines represent disks formed by collisions between icy planets and rocky planets, respectively.}
\label{fig:cummulative}
\end{figure*}

\begin{figure*}
  \begin{center}
\includegraphics[scale=1.1]{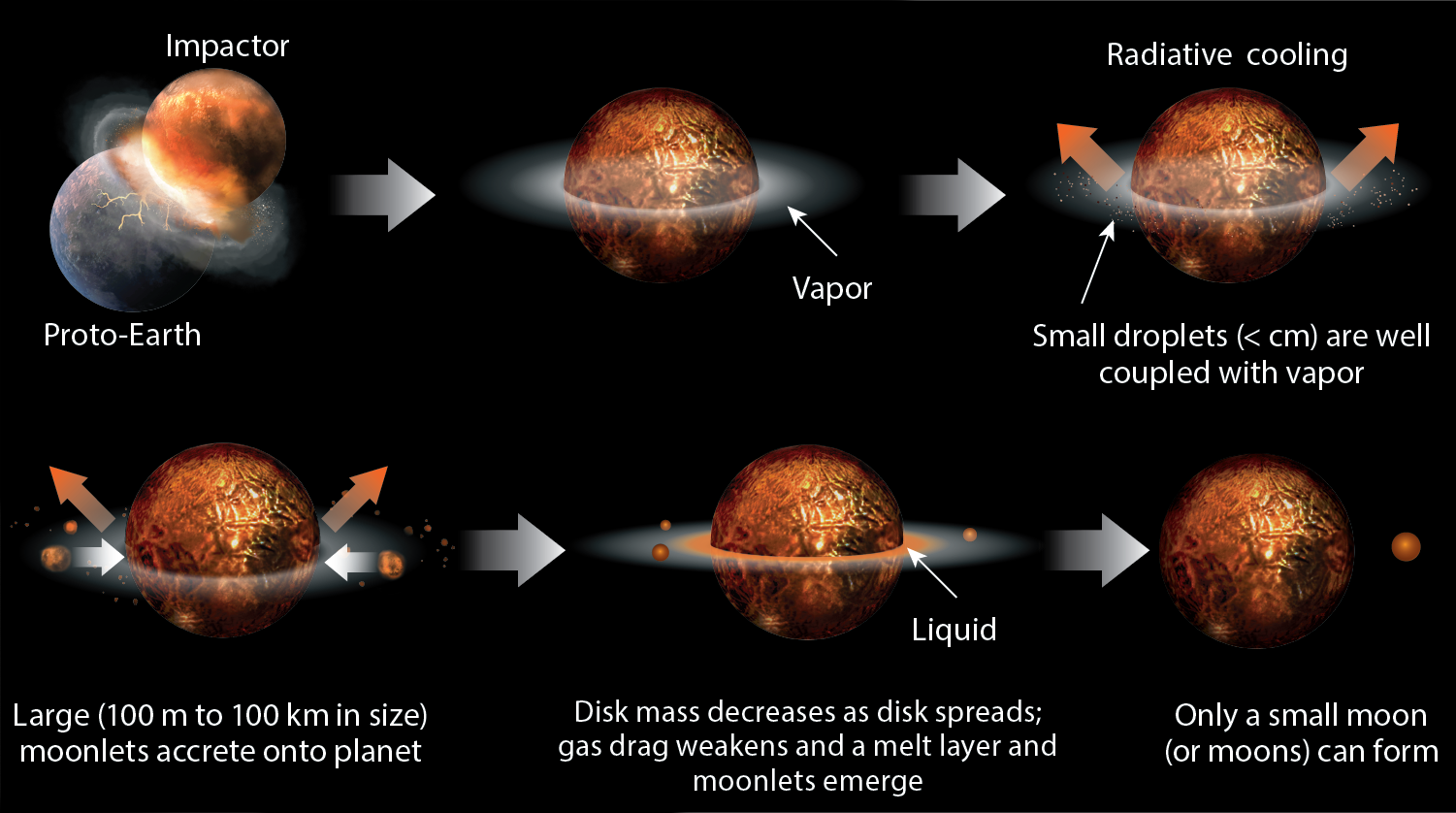}
  \end{center}
  \caption{Schematic view of the Moon-formation from an initially vapor-rich disk.}
\label{fig:bigpicture}
\end{figure*}

Figure \ref{fig:spacetime}A shows four examples of space time plots of our 2D simulations, where (1) $(\tau_f, \Delta)= 10^{-3}, 0.3$, (2) $ 10^{-2}, 0.1$, (3) $ 10^{-1}, 0.2$, and (4) $ 10^{-1}, 0.2$. $Z=0.1$ for the four cases. These simulations represent four characteristic regimes. The horizontal axis is $x$ normalized by the scale height $H$ and the vertical axis is the number of orbits. The color shows particle concentration.  
In case (1), $\tau_{\rm f}$ is small, which means that gas and particles are well coupled and this does not lead to filament formation. In case (2), after $\sim 20$ orbits, filaments start forming and these filaments are stable during the rest of the simulations, which indicate streaming instability occurs with this chosen parameter. A radially shearing periodic boundary condition is used in the $x$ direction, and therefore the filament that appears at $x/H=-0.1$ at $\sim100$ orbits reappears at $x/H=0.1$. Two distinct filaments form in this simulation.  In case (3), filaments form, but they are not stable because their radial movements are high due to the high $\Delta$ value. Thus, this is not an ideal parameter space for SI. Similar behaviors have been seen in simulations for the protoplanetary disk with other parameter combinations (e.g. $\Delta=0.05, \tau_{\rm f}=3, Z=0.005$, \citealt{Carreraetal2015}). In case (4), filaments are not as clearly defined as those in case (2), but a coherent filament structure is found after several orbits. Thus, this is also considered as an SI regime.

Figure \ref{fig:spacetime}B shows the summary of our 2D simulations for $Z=0.1$ (top) and for $Z=0.05$ (bottom). 
We use the Kolmogorov–Smirnov test (KS) to identify the extent of particle concentration (see Section \ref{sect:clump}). Concentration is measured by the $p$ value and when the value of $p$ is small ($p<0.1$), streaming instability is likely \citep{Carreraetal2015}. The colorbar indicates the $p$ value. At $Z=0.1$, $p<0.1$ is achieved when $\Delta=0.1$ and $10^{-3}\leq \tau_{\rm f}\leq10^0$. At large $\Delta$ values, radial motions are large and stable filaments do not form, as discussed above. At $Z=0.05$, this trend remains the same, but the streaming instability regime is smaller ($10^{-2}\leq \tau_{\rm f}\leq10^0$) than that of $Z=0.1$. This is because larger $Z$ leads to more particles in the disk and therefore to more filament formation \citep{Carreraetal2015}. Thus, our 2D simulations show that the most filaments form at relatively small $\Delta$ ($\Delta=0.1$) and with both $Z=0.1$ and $Z=0.05$, but the higher $Z$ has more favorable conditions. This general trend is consistent with previous work in the protoplanetary disk (e.g.  \citealt{Carreraetal2015}), which investigate smaller $\Delta$ values for the protoplanetary disk ($\Delta \leq 0.05$). 


\subsection{3D Athena simulations}
\label{sec:results3D}

Now that we identify the parameter space for the streaming instability ($\Delta=0.1$, $Z=0.1$), we perform 3D simulations with self-gravity to estimate the size distributions of SI-induced moonlets.  Figure \ref{fig:3D} shows snapshots of one of our 3D simulations ($\tau_f=1, \tilde G=0.5898, Z=0.1$ and $\Delta=0.1$) at four different times ($t=0.32, 2.87, 3.39, 3.55$), where $t$ is the number of orbits. The self-gravity is not initially included and is turned on at $t=3.18$. This is a general practice to avoid artificial clumping before streaming instability takes place \citep{Simonetal2016}. 
The color shows the particle density. The circles indicate the Hill spheres of self-bound clumps, which are identified using PLAN \citep{Lietal2019}. At $t=0.32$, no clump is identified, but by $t=2.87$, the streaming instability fully develops and reaches its steady state. After self gravity is turned on at $t=3.18$, self-gravitating clumps form right away ($t=3.39$). This general behavior is similar to previous work on the streaming instability in the protoplanetary disk \citep{Abodetal2019}, but the time it takes to form clumps by streaming instability is shorter, 
probably because of the large $Z$ value \citep{Simonetal2022} compared to those of the protoplanetary disk.  

Figure \ref{fig:cummulative}A shows the cumulative distribution of moonlet mass $M_p$, normalized by the characteristic self-gravitating mass $M_G$, defined as \citep{Abodetal2019},
\begin{equation} 
\label{mass_estimate}
M_G  = \pi \left( \frac{\lambda_G}{2} \right)^2 \Sigma_p = 4 \pi^5 \frac{G^2 \Sigma_p^3}{\Omega^4}\\
  = \frac{\sqrt{2}}{2}\pi^{9/2}Z^3 \tilde{G}^2 (\rho_{\rm g,0} H^3),
\end{equation}
where $\lambda_G$ is an instability wavelength, which  originates from the Toomre dispersion relation, equating the tidal and gravitational forces \citep{Abodetal2019}. 
The parameters are $\tau_f=0.1, 1$, $\tilde G=0.1788, 0.5898$ and $Z=0.1, 0.3$. 
For the $\tilde{G}=0.1787$ cases, $M_G=5.19\times 10^{18}$ kg and for the $\tilde{G}=0.5898$ cases, $M_G=2.17\times 10^{20}$ kg. For all the 3D simulations, $\Delta=0.1$ are assumed. 
 The solid lines indicate the moonlet mass distribution shortly after the onset of the self gravity ($t=3.66$ for $\tilde{G}=0.1788,\tau_{\rm f}=0.1$ and $t=3.34$ for all the other cases). The dashed lines indicate the same parameter after an additional time ($t+\Delta t$, where $\Delta t=1/\Omega =$ 0.16 orbit except the $Z=0.3$ case where $\Delta t=\frac{1}{2\Omega}$. The reason of the shorter $\Delta t$ for the higher $Z$ case is that the streaming instability develops faster for higher $Z$ values, \citealt{Simonetal2022}).
The maximum $M_p/M_G=0.254$ ($\tilde G=0.1788, \tau_f=1$ at $t=3.50$). This is broadly consistent with previous studies that suggest that the maximum clump masses formed by the streaming instability are characterized by $M_G$, ranging $10^{-1}-10^1 M_G$ \citep{Abodetal2019, Lietal2019}. Our result lies on the lower end of this mass range. This indicates that the moonlet mass distribution based on our numerical simulations is consistent with the analytical mass model generated for the protoplanetary disk.  Thus, this also indicates that the processes of streaming instability are similar between the Moon-forming disk and protoplanetary disk despite the different input parameters. We also find that higher $Z$ does not necessarily lead to higher $M_p$.



Here we consider the best case scenario of forming a large moonlet. Assuming that the density of the clump is $3000$ kg m$^{-3}$, then $M_G = 2.17 \times 10^{20}$ kg is equivalent to 258 km in radius and 0.254 $M_G = 163$ km. Based on equation (\ref{eq:vr_main}),
the residence time of 258 km and 163 km moonlets are 92 and 58 days, respectively, assuming $v_{ r,{\rm g}}=0$. These timescales are much shorter than the Moon-formation timescale (10s-100 years, \citealt{ThompsonStevenson1988}). Thus, even though the streaming instability can occur in an initially vapor-rich Moon-forming disks, it does not help increasing the residence time of moonlets. 

\subsection{Exomoon formation by streaming instability}
\label{sec:resultsexomoon}
The streaming instability likely occurs in impact-induced moon-forming disks in extrasolar systems. The pressure gradient parameter $\eta$ ($\sim 0.02-0.06$) is similar regardless of the composition of the disk \citep{Nakajimaetal2022}. Figure \ref{fig:cummulative}B shows the disk residence time for a clump with mass $M_G$ in a moon-forming disk as a function of the planetary mass ($M_{\rm planet} = 1-6M_\oplus$). ``Rocky Planet'' corresponds to disks formed by a collision between terrestrial planets while ``Icy Planet'' indicates disks formed by a collision between icy planets whose mantles are made of water ice (70 wt\%) and cores are made of iron (30 wt\%). This is produced by calculating $r/v_r$ (see Equation \ref{eq:vr_main}), where for rocky planets we use $\rho_{g,0} = 40.01$ kg m$^{-3}$, $T=5200$ K, $\rho_p=3000$ kg m$^{-3}$, $r=3 R_{\rm planet}$ where $R_{\rm planet}$ is the planetary radius, and for icy planets we use $\rho_{g,0} = 10.0$ kg m$^{-3}$, $T=2000$ K, $\rho_p=1000$ kg m$^{-3}$, $r=3 R_{\rm planet}$. These are the temperatures when the impact-induced disks reach complete vaporization based on impact simulations  \citep{Nakajimaetal2022}. Higher temperature is possible, but the disk would need to cool down to this temperature to form particles (dust), and therefore this is the most relevant temperature to assess the effect of streaming instability. The thermal state of the disk formed by icy planetary collisions are estimated based on an equation of state of water \citep{SenftStewart2008}. 
Additionally, $R_{\rm planet} = R_\oplus(M_{\rm planet}/M_\oplus)^{1/3.7}$ for rocky planets and $R_{\rm planet} = 1.2R_\oplus(M_{\rm planet}/M_\oplus)^{1/3}$ for icy planets are assumed \citep{Kippingetal2013, Mordasinietal2012}. At $M_{\rm planet}=1-6 M_\oplus$, these parameters make $\tilde G=0.42-0.67$ for rocky planets and $\tilde G=0.25$ for icy planets, which are similar to the ranges covered in our hydrodynamic simulations (Section \ref{sec:modelparameters}).

In both icy and rocky planet cases, the disk residence time is a few 10s of days to several months, which are short compared to the satellite formation timescale (10s-100s years, \citealt{Nakajimaetal2022}). 
Another effect is that as the disk cools, $\rho_{\rm g,0}$ and $H$ decrease, which means that $M_G$ decreases over time in both the icy and rocky disks. This means that SI-induced clumps tend to become smaller as time progresses. These effects indicate that the streaming instability likely plays a limited role in impact-induced moon-forming disks.


\section{Discussion}
\subsection{Streaming instability in the Moon-forming disk}
\label{sec:SI_Moon}

Figure \ref{fig:bigpicture} shows a schematic view of our hypothesis. An energetic impact would generate a vapor-rich disk (the disk is made of silicate vapor for the Moon or rocky planets and water vapor for icy planets). Over time, the disk cools by radiation and small droplets ($<$ cm) emerge. These small droplets would grow by accretion and by streaming instability. However, once these moonlets reach 100 m - 100 km in size, gas drag from the vapor is so strong that they fall onto the planet on a short time scale (days-weeks). This continues to occur until the vapor mass fraction of the disk decreases so that the gas drag effect is no longer strong. Once this condition is reached, a liquid layer emerges and moonlets can stay in the disk. However, by this time a significant disk mass could have been lost ($>$ 80 wt \%) \citep{Idaetal2020, Nakajimaetal2022}. For this reason, only a small moon (or moons) could form from an initially vapor-rich disk. Thus, we suggest that an initially vapor-rich disk is not suitable for forming our Moon, and our result supports the hypothesis that the Moon formed from an initially vapor-poor disk, including the canonical model where the proto-Earth was hit by a Mars-sized impactor \citep{CanupAsphaug2001}.


The isotopic composition of the Moon may constrain whether streaming instability played a role during the Moon formation. Some may argue that the observational fact that the Moon is depleted in volatiles may be explained by accreting moonlets formed by streaming instability before volatiles accreted onto the Moon. However, this process would make the Moon isotopically light if these isotopes experience kinetic fractionation \citep{Dauphasetal2015, Dauphasetal2022}, which is inconsistent with the observation of enrichment of heavy isotopes in the Moon (such as K, \citealt{WangJacobsen2016}). The lunar isotopes would be heavier if they experience equilibrium fractionation, but the equilibrium fractionation would not produce the observed isotopic fractionations at the high disk temperature condition. Alternatively, some volatiles could be lost from the lunar magma ocean under an equilibrium condition \citep{Charnozetal2021}, but the efficiency of volatile loss is not fully known \citep{Dauphasetal2022}.


\subsection{Streaming instability in general impact-induced disks}
Our model supports the previous work that suggests that relatively large rocky ($>6 M_\oplus$, $>1.6 R_\oplus$) and icy ($>1 M_\oplus$, $>1.3 R_\oplus$) planets cannot form impact-induced moons that are large compared to the host planets \citep{Nakajimaetal2022}. Larger planets than those thresholds generate completely vapor disks, because the kinetic energy involved in an impact scales with the planetary mass. Thus, these large planets are not capable of forming moons that are large compared to their host planets. Moons can form by mechanisms other than impacts, such as formation in a circumplanetary disk and gravitational capture, but these moons tend to be small compared to the sizes of their host planets (the predicted moon to planet mass ratio is $\sim 10^{-4}$ for moons formed by circumplanetary disk \citealt{CanupWard2006} and gravitationally captured moons are small in the solar system, \citealt{AgnorHamilton2006}). Thus, fractionally large moons compared to the host planet sizes, which are observationally favorable, likely form by impact.
So far, no exomoon has been confirmed despite extensive searches, but future observations, especially with James Webb Space Telescope \citep{CasseseKipping2022}, may be able to find exomoons and test this theoretical hypothesis.

\subsection{Comparison between the Moon-forming disk and protoplanetary disk}

It is certainly intriguing that the streaming instability is able to solve the gas drag problem in the protoplanetary disk, but not in the Moon-forming disk. In both scenarios, the clump sizes formed by the streaming instability happen to be similar ($M_G \sim$100 km, see \citealt{Johansenetal2015}, for the protoplanetary disk and $\sim$100 km for the protolunar disk, respectively). This size is $\sim10^5$ times larger than the size of the particle (0.52-1 m)\citep{Adachietal1976} that experiences the strongest gas drag ($\tau_{\rm f}=1$ and $v_r \sim -\eta v_{\rm K}$). Therefore, once particles become $\sim 100$ km-sized planetesimals by streaming instability, the radial velocity of the planetesimals decrease drastically ($v_r=-\frac{2\eta v_{\rm K}}{\tau_{\rm f}}$ for large ${\tau_{\rm f}}$, see Equation \ref{eq1}), which helps planetesimal growth. In contrast, the largest moonlet size (100 km) formed by streaming instability in the Moon-forming disk is only 50 times larger than the particle size that experiences the strongest gas drag (2km). This does not result in a large $\tau_f$ change or $v_r$ change, and therefore streaming instability does not effectively help moon formation. Therefore, streaming instability is an effective mechanism to skip the ``1 m barrier'' in the protoplanetary disk, whereas it is not an effective mechanism to skip the ``km barrier'' in the Moon-forming disk.


\subsection{Streaming instability in the circumplanetary disk around a gas giant}
\label{ref:gasgiants}
The streaming instability may occur during moon formation in various satellite systems around gas giants. The Galilean moons around Jupiter and Titan around Saturn likely formed from their circumplanetary disks. In these disks, the particle-to-gas ratio $Z$ is typically $10^{-4}-10^{-2}$, which was thought to be too small for streaming instability to occur \citep{Carreraetal2015, Yangetal2017, Shibaikeetal2017}. However, recent SI calculations show that SI can occur at as low as $Z=4 \times 10^{-3}$ (at $\Delta=0.05$ and $\tau_{\rm f}=0.3$, \citealt{LiYoudin2021}, and the required $Z$ depends on the disk conditions, e.g. \citealt{Carreraetal2015, Yangetal2017, SekiyaOnishi2018}) which may indicate that SI can be potentially important in circumplanetary disks around gas giants if the dust-gas ratio of the disk is relatively high.  The velocity difference between the gas and particles $v_{\rm rel}$ is \citep{CanupWard2002} 
\begin{equation}
 v_{\rm rel} = \eta v_{\rm k} \sim c_{\rm s} \left( \frac{c_{\rm s}}{r\Omega} \right).
    \label{eq:pressuregradient}
\end{equation}
 Assuming $\frac{c_{\rm s}}{r\Omega}= \frac{c_{\rm s}}{v_k}\sim0.1$ \citep{CanupWard2002}, this yields $\eta \sim 0.01$ and $\Delta=\frac{\eta v_k}{c_{\rm s}} \sim  0.1 $. If $Z$ is sufficiently large (at least $Z>4 \times 10^{-3}$   \citealt{LiYoudin2021}), it is possible that the streaming instability takes place in a circumplanetary disk around a gas giant. A rough estimate for the size of a moonlet in a circumplanetary disk is $M_{\rm G}=1.4\times10^{16}$ kg (Equation \ref{mass_estimate}), which is $\sim 10.3$ km in radius if it is a rocky moonlet. Here, we are assuming $Z=10^{-2}$, $\Sigma_p=3 \times 10^4$ kg m$^{-2}$, $c_{\rm s}=1000$ m s$^{-1}$, $H\sim c_{\rm s}/\Omega$, $\rho_{\rm g} = \Sigma_{\rm g}/(H \sqrt{2\pi})$ \citep{CanupWard2002}. These moonlets are relatively small compared to large moons around gas giants (e.g. the Galiean moon masses range $10^{22}-10^{23}$ kg) and it is unclear if they would significantly impact these moon-formation process. Further research is needed to understand its effect on the satellite formation in a circumplanetary disk around gas giants.

\subsection{Model limitations}
There are several model limitations that need to be addressed in our future work. First, the effect of the Roche radius ($a_R \sim 3 R_{\oplus}$) needs to be taken into account to understand the Moon formation. An inspiraling moonlet would not directly reach the Earth, but would be tidally disrupted near the Roche radius, where it then might be incorporated into an accretion disk (e.g., \citealt{SalmonCanup2014}). Secondly, our model presented here does not take into account the evolution of the disk in detail, which is important to identify the final mass and composition of the resulting moon. As the disk spreads out, it is likely that $\Delta$ decreases, which would slow down the radial drift of moonlets. This gas drag effect would disappear once the local vapor condenses, which would occur at the outer part of the disk first, since that part of the disk can cool efficiently due to its large surface area.  Given that such moonlets that directly form from the disk would not have time to lose volatiles in the disk phase and therefore this model requires that the Moon lost its volatiles before or after the disk phase, for example during the lunar magma ocean phase \citep{Charnozetal2021} (see further discussion in Section \ref{sec:SI_Moon}).  
Nevertheless, such scenario may be possible in moon-forming disks in the solar and extrasolar systems. The effects of the Roche radius and disk evolution will be addressed in our future work.

\section{Conclusion}
In conclusion, we show that the streaming instability can form self-gravitating clumps ($\sim10^2$ km) in a vapor-rich moon-forming disk generated by a giant impact, but the sizes of the clumps it generates are not large enough  to avoid inspiraling due to the strong gas drag. This is a major difference from the protoplanetary disk, where the streaming instability can efficiently form large clumps to avoid the strong gas drag effect. As a result, growing moonlets in an initially vapor-rich moon-forming disk continue to fall onto the planet once they reach the sizes of 100 m - 100 km. These moonlets could grow further once the disk cools enough and the vapor mass fraction of the disk becomes small. However, by this time a significant amount of the disk mass is lost, and the remaining disk could make only a small moon. 
This result is applicable to impact-induced moon-forming disks in the solar system and beyond; we find that the streaming instability is not an efficient mechanism to form a large moon from an impact-induced vapor-rich disks in general. As a result, we support previous work that suggests that fractionally-large moons compared to their host planets form from vapor-poor disks; the ideal planetary radii that host fractionally large moons are $1.3-1.6 R_\oplus$ \citep{Nakajimaetal2022} given that rocky or icy planets larger than these sizes would likely produce completely vapor disks, which are not capable of forming large moons \citep{Nakajimaetal2022}. The streaming instability may take place in circumplanetary disks, but their effect on the moon-formation process needs further investigation.

\section*{Acknowledgments}
We thank code developers of Athena \citep{Stoneetal2008, BaiStone2010a, Simonetal2016} and PLAN \citep{Lietal2019}. We appreciate discussion with Shigeru Ida and Scott D. Hull. J.A. was partially supported by the Research Experience for Undergraduates (REU) Program, a National Science Foundation (NSF), under Grand No. PHY-1757062.
M.N. was supported in part by the National Aeronautics and Space Administration (NASA) grant numbers 80NSSC19K0514, 80NSSC21K1184, and 80NSSC22K0107. Partial funding for M.N. was also provided by NSF EAR-2237730 as well as the Center for Matter at Atomic Pressures (CMAP), an NSF Physics Frontier Center, under Award PHY-2020249. Any opinions, findings, conclusions or recommendations expressed in this material are those of the authors and do not necessarily reflect those of the National Science Foundation. M.N. was also supported in part by the Alfred P. Sloan Foundation under grant G202114194.

\bibliography{sample631.bbl}

\begin{thebibliography}{}
\expandafter\ifx\csname natexlab\endcsname\relax\def\natexlab#1{#1}\fi
\providecommand{\url}[1]{\href{#1}{#1}}
\providecommand{\dodoi}[1]{doi:~\href{http://doi.org/#1}{\nolinkurl{#1}}}
\providecommand{\doeprint}[1]{\href{http://ascl.net/#1}{\nolinkurl{http://ascl.net/#1}}}
\providecommand{\doarXiv}[1]{\href{https://arxiv.org/abs/#1}{\nolinkurl{https://arxiv.org/abs/#1}}}

\bibitem[{Abod {et~al.}(2019)Abod, Simon, Li, Armitage, Youdin, \& Kretke}]{Abodetal2019}
Abod, C.~P., Simon, J.~B., Li, R., {et~al.} 2019, The Astronomical Journal, 883, 192, \dodoi{10.3847/1538-4357/ab40a3}

\bibitem[{Adachi {et~al.}(1976)Adachi, Hayashi, \& Nakazawa}]{Adachietal1976}
Adachi, I., Hayashi, C., \& Nakazawa, K. 1976, Progress of Theoretical Physics, 56, 1756, \dodoi{10.1143/PTP.56.1756}

\bibitem[{Agnor \& Hamilton(2006)}]{AgnorHamilton2006}
Agnor, C.~B., \& Hamilton, D.~P. 2006, Nature, 441, 192, \dodoi{10.1038/nature04792}

\bibitem[{Armitage(2010)}]{Armitage2010}
Armitage, P.~J. 2010, Cambridge University Press

\bibitem[{Armytage {et~al.}(2012)Armytage, Georg, Williams, \& Halliday}]{Armytageetal2012}
Armytage, R. M.~G., Georg, R.~B., Williams, H.~M., \& Halliday, A.~N. 2012, Geochim. Cosmochia. Ac., 77, 504, \dodoi{10.1016/j.gca.2011.10.032}

\bibitem[{Asphaug {et~al.}(2021)Asphaug, Emsenhuber, Cambioni, Gabriel, \& Schwartz}]{Asphaugetal2021}
Asphaug, E., Emsenhuber, A., Cambioni, S., Gabriel, T. S.~J., \& Schwartz, S.~R. 2021, The Planetary Science Journal, 2, 200, \dodoi{10.3847/PSJ/ac19b2}

\bibitem[{Bai \& Stone(2010{\natexlab{a}})}]{BaiStone2010a}
Bai, X.-N., \& Stone, J.~M. 2010{\natexlab{a}}, The Astrophysical Journal, 722, 1437, \dodoi{10.1088/0004-637x/722/2/1437}

\bibitem[{Bai \& Stone(2010{\natexlab{b}})}]{BaiStone2010b}
---. 2010{\natexlab{b}}, The Astrophysical Journal Supplement Series, 190, 297, \dodoi{10.1088/0067-0049/190/2/297}

\bibitem[{Bonomo {et~al.}(2019)Bonomo, Zeng, Damasso, Leinhardt, Justesen, Lopez, Lund, Malavolta, Silva~Aguirre, Buchhave, Corsaro, Denman, Lopez-Morales, Mills, Mortier, Rice, Sozzetti, Vanderburg, Affer, Arentoft, Benbakoura, Bouchy, Christensen-Dalsgaard, Collier~Cameron, Cosentino, Dressing, Dumusque, Figueira, Fiorenzano, García, Handberg, Harutyunyan, Johnson, Kjeldsen, Latham, Lovis, Lundkvist, Mathur, Mayor, Micela, Molinari, Motalebi, Nascimbeni, Nava, Pepe, Phillips, Piotto, Poretti, Sasselov, Ségransan, Udry, \& Watson}]{Bonomoetal2019}
Bonomo, A.~S., Zeng, L., Damasso, M., {et~al.} 2019, Nature Astronomy, 3, 416, \dodoi{10.1038/s41550-018-0684-9}

\bibitem[{Bouhifd {et~al.}(2020)Bouhifd, Jephcoat, Porcelli, Kelley, \& Marty}]{Bouhifdetal2020}
Bouhifd, A.~M., Jephcoat, A.~P., Porcelli, D., Kelley, S.~P., \& Marty, B. 2020, Geochemical Perspectives Letters, 15, 15, \dodoi{10.7185/geochemlet.2028}

\bibitem[{Cameron \& Ward(1976)}]{CameronWard1976}
Cameron, A. G.~W., \& Ward, W.~R. 1976, Lunar Planet. Sci. VII, 120

\bibitem[{Cano {et~al.}(2020)Cano, Sharp, \& Shearer}]{Canoetal2020}
Cano, E.~J., Sharp, Z.~D., \& Shearer, C.~K. 2020, Nature Geoscience, 13, 270, \dodoi{10.1038/s41561-020-0550-0}

\bibitem[{Canup(2004)}]{Canup2004}
Canup, R.~M. 2004, Icarus, 168, 433, \dodoi{10.1016/j.icarus.2003.09.028}

\bibitem[{Canup(2005)}]{Canup2005}
---. 2005, Science, 307, 546, \dodoi{10.1126/science.1106818}

\bibitem[{Canup(2012)}]{Canup2012}
---. 2012, Science, 338, 1052, \dodoi{10.1126/science.1226073}

\bibitem[{Canup \& Asphaug(2001)}]{CanupAsphaug2001}
Canup, R.~M., \& Asphaug, E. 2001, Nature, 412, 708, \dodoi{10.1038/35089010}

\bibitem[{Canup {et~al.}(2015)Canup, Visscher, Salmon, \& Fegley}]{Canupetal2015}
Canup, R.~M., Visscher, C., Salmon, J., \& Fegley, B. 2015, Nature Geoscience, 8, 918, \dodoi{10.1038/ngeo2574}

\bibitem[{Canup \& Ward(2002)}]{CanupWard2002}
Canup, R.~M., \& Ward, W.~R. 2002, The Astronomical Journal, 124, 3404, \dodoi{10.1086/344684}

\bibitem[{Canup \& Ward(2006)}]{CanupWard2006}
---. 2006, Nature, 441, 834, \dodoi{10.1038/nature04860}

\bibitem[{Canup {et~al.}(2023)Canup, Righter, Dauphas, Pahlevan, Ćuk, Lock, Stewart, Salmon, Rufu, Nakajima, \& Magna}]{Canupetal2023}
Canup, R.~M., Righter, K., Dauphas, N., {et~al.} 2023, Reviews in Mineralogy and Geochemistry, 89, 53, \dodoi{10.2138/rmg.2023.89.02}

\bibitem[{Carrera {et~al.}(2015)Carrera, Johansen, \& Davies}]{Carreraetal2015}
Carrera, D., Johansen, A., \& Davies, M.~B. 2015, Astronomy \& Astrophysics, 579, A43, \dodoi{10.1051/0004-6361/201425120}

\bibitem[{Cassese \& Kipping(2022)}]{CasseseKipping2022}
Cassese, B., \& Kipping, D. 2022, Monthly Notices of the Royal Astronomical Society, 516, 3701, \dodoi{10.1093/mnras/stac2090}

\bibitem[{Charnoz \& Michaut(2015)}]{CharnozMichaut2015}
Charnoz, S., \& Michaut, C. 2015, Icarus, 260, 440, \dodoi{10.1016/j.icarus.2015.07.018}

\bibitem[{Charnoz {et~al.}(2021)Charnoz, Sossi, Lee, Siebert, Hyodo, Allibert, Pignatale, Landeau, Oza, \& Moynier}]{Charnozetal2021}
Charnoz, S., Sossi, P.~A., Lee, Y.-N., {et~al.} 2021, Icarus, 364, 114451, \dodoi{10.1016/j.icarus.2021.114451}

\bibitem[{Chiang \& Youdin(2010)}]{ChiangYoudin2010}
Chiang, E., \& Youdin, A.~N. 2010, Annual Review of Earth and Planetary Sciences, 38, 493, \dodoi{10.1146/annurev-earth-040809-152513}

\bibitem[{Colella(1990)}]{Colella1990}
Colella, P. 1990, Journal of Computational Physics, 87, 171, \dodoi{10.1016/0021-9991(90)90233-Q}

\bibitem[{Colella \& Woodward(1984)}]{ColellaWoodward1984}
Colella, P., \& Woodward, P.~R. 1984, Journal of Computational Physics, 54, 174, \dodoi{10.1016/0021-9991(84)90143-8}

\bibitem[{Craddock(2011)}]{Craddock2011}
Craddock, R.~A. 2011, Icarus, 211, 1150, \dodoi{10.1016/j.icarus.2010.10.023}

\bibitem[{C\'{u}k {et~al.}(2016)C\'{u}k, Hamilton, Lock, \& Stewart}]{Cuketal2016}
C\'{u}k, M., Hamilton, D.~P., Lock, S.~J., \& Stewart, S.~T. 2016, Nature, 539, 402, \dodoi{10.1038/nature19846}

\bibitem[{C\'{u}k \& Stewart(2012)}]{CukStewart2012}
C\'{u}k, M., \& Stewart, S.~T. 2012, Science, 338, 1047, \dodoi{10.1126/science.1225542}

\bibitem[{Dauphas(2017)}]{Dauphas2017}
Dauphas, N. 2017, Nature, 541, 521, \dodoi{10.1038/nature20830}

\bibitem[{Dauphas {et~al.}(2015)Dauphas, Poitrasson, Burkhardt, Kobayashi, \& Kurosawa}]{Dauphasetal2015}
Dauphas, N., Poitrasson, F., Burkhardt, C., Kobayashi, H., \& Kurosawa, K. 2015, Earth and Planetary Science Letters, 427, 236, \dodoi{10.1016/j.epsl.2015.07.008}

\bibitem[{Dauphas {et~al.}(2022)Dauphas, Nie, Blanchard, Zhang, Zeng, Hu, Meheut, Visscher, Canup, \& Hopp}]{Dauphasetal2022}
Dauphas, N., Nie, N.~X., Blanchard, M., {et~al.} 2022, The Planetary Science Journal, 3, 29, \dodoi{10.3847/PSJ/ac2e09}

\bibitem[{Halliday \& Canup(2022)}]{HallidayCanup2022}
Halliday, A.~N., \& Canup, R.~M. 2022, Nature Reviews Earth \& Environment, 4, 19, \dodoi{10.1038/s43017-022-00370-0}

\bibitem[{Halliday(2004)}]{Halliday2004}
Halliday, N.~H. 2004, Nature, 427, 505

\bibitem[{Hartmann \& Davis(1975)}]{HartmannDavis1975}
Hartmann, W.~K., \& Davis, D.~R. 1975, Icarus, 24, 504, \dodoi{10.1016/0019-1035(75)90070-6}

\bibitem[{Hosono {et~al.}(2019)Hosono, Karato, Makino, \& Saitoh}]{Hosonoetal2019}
Hosono, N., Karato, S.-i., Makino, J., \& Saitoh, T.~R. 2019, Nature Geoscience, 12, 418, \dodoi{10.1038/s41561-019-0354-2}

\bibitem[{Hull {et~al.}(2024)Hull, Nakajima, Hosono, Canup, \& Gassmöller}]{Hulletal2024}
Hull, S.~D., Nakajima, M., Hosono, N., Canup, R.~M., \& Gassmöller, R. 2024, The Planetary Science Journal, 5, 9, \dodoi{10.3847/PSJ/ad02f7}

\bibitem[{Ida {et~al.}(2020)Ida, Ueta, Sasaki, \& Ishizawa}]{Idaetal2020}
Ida, S., Ueta, S., Sasaki, T., \& Ishizawa, Y. 2020, Nature Astronomy, 4, 880 \dodoi{10.1038/s41550-020-1049-8}

\bibitem[{Johansen {et~al.}(2015)Johansen, Mac~Low, Lacerda, \& Bizzarro}]{Johansenetal2015}
Johansen, A., Mac~Low, M.~M., Lacerda, P., \& Bizzarro, M. 2015, Science Advances, 1, :e1500109

\bibitem[{Johansen {et~al.}(2007)Johansen, Oishi, Mac~Low, Klahr, Henning, \& Youdin}]{Johansenetal2007}
Johansen, A., Oishi, J.~S., Mac~Low, M.~M., {et~al.} 2007, Nature, 448, 1022, \dodoi{10.1038/nature06086}

\bibitem[{Kegerreis {et~al.}(2022)Kegerreis, Ruiz-Bonilla, Eke, Massey, Sandnes, \& Teodoro}]{Kegerreisetal2022}
Kegerreis, J.~A., Ruiz-Bonilla, S., Eke, V.~R., {et~al.} 2022, The Astrophysical Journal Letters, 937, 40, \dodoi{10.3847/2041-8213/ac8d96}

\bibitem[{Kenworthy {et~al.}(2023)Kenworthy, Lock, Kennedy, van Capelleveen, Mamajek, Carone, Hambsch, Masiero, Mainzer, Kirkpatrick, Gomez, Leinhardt, Dou, Tanna, Sainio, Barker, Charbonnel, Garde, Le~Du, Mulato, Petit, \& Rizzo~Smith}]{Kenworthyetal2023}
Kenworthy, M., Lock, S., Kennedy, G., {et~al.} 2023, Nature, 622, 251, \dodoi{10.1038/s41586-023-06573-9}

\bibitem[{Kipping {et~al.}(2013)Kipping, Forgan, Hartman, Nesvorný, Bakos, Schmitt, \& Buchhave}]{Kippingetal2013}
Kipping, D.~M., Forgan, D., Hartman, J., {et~al.} 2013, The Astrophysical Journal, 777, 134, \dodoi{10.1088/0004-637x/777/2/134}

\bibitem[{Krapp {et~al.}(2019)Krapp, Benítez-Llambay, Gressel, \& Pessah}]{Krappetal2019}
Krapp, L., Benítez-Llambay, P., Gressel, O., \& Pessah, M.~E. 2019, The Astrophysical Journal, 878, 30, \dodoi{10.3847/2041-8213/ab2596}

\bibitem[{Kruijer {et~al.}(2021)Kruijer, Archer, \& Kleine}]{Krujiferetal2021}
Kruijer, T.~S., Archer, G.~J., \& Kleine, T. 2021, Nature Geoscience, 14, 714, \dodoi{10.1038/s41561-021-00820-2}

\bibitem[{Kruijer {et~al.}(2015)Kruijer, Kleine, Fischer-Godde, \& Sprung}]{Kruijeretal2015}
Kruijer, T.~S., Kleine, T., Fischer-Godde, M., \& Sprung, P. 2015, Nature, 520, 534, \dodoi{10.1038/nature14360}

\bibitem[{Li \& Youdin(2021)}]{LiYoudin2021}
Li, R., \& Youdin, A.~N. 2021, The Astrophysical Journal, 919, 107, \dodoi{10.3847/1538-4357/ac0e9f}

\bibitem[{Li {et~al.}(2019)Li, Youdin, \& Simon}]{Lietal2019}
Li, R., Youdin, A.~N., \& Simon, J.~B. 2019, The Astrophysical Journal, 885, 69, \dodoi{10.3847/1538-4357/ab480d}

\bibitem[{Lock {et~al.}(2018)Lock, Stewart, Petaev, Leinhardt, Mace, Jacobsen, \& Cuk}]{Locketal2018}
Lock, S.~J., Stewart, S.~T., Petaev, M.~I., {et~al.} 2018, Journal of Geophysical Research: Planets, 123, 910, \dodoi{10.1002/2017je005333}

\bibitem[{Machida \& Abe(2004)}]{MachidaAbe2004}
Machida, R., \& Abe, Y. 2004, Astrophys. J., 617, 633, \dodoi{10.1086/425209}

\bibitem[{Melosh(2007)}]{Melosh2007}
Melosh, H.~J. 2007, Meteorit. Planet. Sci., 42, 2079, \dodoi{10.1111/j.1945-5100.2007.tb01009.x}

\bibitem[{Meng {et~al.}(2014)Meng, Su, Rieke, Stevenson, Plavchan, Rujopakarn, Lisse, Poshyachinda, \& Richart}]{Mengetal2014}
Meng, H. Y.~A., Su, K. Y.~L., Rieke, G.~H., {et~al.} 2014, Science, 345, 1032, \dodoi{10.1126/science.1255153}

\bibitem[{Mordasini {et~al.}(2012)Mordasini, Alibert, Georgy, Dittkrist, Klahr, \& Henning}]{Mordasinietal2012}
Mordasini, C., Alibert, Y., Georgy, C., {et~al.} 2012, Astronomy \& Astrophysics, 547, 112, \dodoi{10.1051/0004-6361/201118464}

\bibitem[{Mullen \& Gammie(2020)}]{MullenGammie2020}
Mullen, P.~D., \& Gammie, C.~F. 2020, The Astrophysical Journal, 903, L15, \dodoi{10.3847/2041-8213/abbffd}

\bibitem[{Nakajima {et~al.}(2022)Nakajima, Genda, Asphaug, \& Ida}]{Nakajimaetal2022}
Nakajima, M., Genda, H., Asphaug, E., \& Ida, S. 2022, Nat Commun, 13, 568, \dodoi{10.1038/s41467-022-28063-8}

\bibitem[{Nakajima \& Stevenson(2014)}]{NakajimaStevenson2014}
Nakajima, M., \& Stevenson, D.~J. 2014, Icarus, 233, 259

\bibitem[{Nakajima \& Stevenson(2015)}]{NakajimaStevenson2015}
---. 2015, Earth and Planetary Science Letters, 427, 286, \dodoi{10.1016/j.epsl.2015.06.023}

\bibitem[{Nakajima \& Stevenson(2018)}]{NakajimaStevenson2018}
---. 2018, Earth and Planetary Science Letters, 487, 117, \dodoi{10.1016/j.epsl.2018.01.026}

\bibitem[{Nie \& Dauphas(2019)}]{NieDauphas2019}
Nie, N.~X., \& Dauphas, N. 2019, The Astrophysical Journal, 884, 48, \dodoi{10.3847/2041-8213/ab4a16}

\bibitem[{Nielsen {et~al.}(2021)Nielsen, Bekaert, \& Auro}]{Nielsenetal2021}
Nielsen, S.~G., Bekaert, D.~V., \& Auro, M. 2021, Nat Commun, 12, 1817, \dodoi{10.1038/s41467-021-22155-7}

\bibitem[{Pahlevan \& Stevenson(2007)}]{PahlevanStevenson2007}
Pahlevan, K., \& Stevenson, D.~J. 2007, Earth and Planetary Science Letters, 262, 438, \dodoi{10.1016/j.epsl.2007.07.055}

\bibitem[{Rufu {et~al.}(2017)Rufu, Aharonson, \& Perets}]{Rufuetal2017}
Rufu, R., Aharonson, O., \& Perets, H.~B. 2017, Nature Geoscience, 10, 89, \dodoi{10.1038/ngeo2866}

\bibitem[{Rufu \& Canup(2020)}]{RufuCanup2020}
Rufu, R., \& Canup, R.~M. 2020, Journal of Geophysical Research: Planets, 125, e2019JE006312, \dodoi{10.1029/2019je006312}

\bibitem[{Salmon \& Canup(2012)}]{SalmonCanup2012}
Salmon, J., \& Canup, R.~M. 2012, Astrophys. J., 760, 83, \dodoi{10.1088/0004-637x/760/1/83}

\bibitem[{Salmon \& Canup(2014)}]{SalmonCanup2014}
---. 2014, Philos Trans A Math Phys Eng Sci, 372, 20130256, \dodoi{10.1098/rsta.2013.0256}

\bibitem[{Sekiya \& Onishi(2018)}]{SekiyaOnishi2018}
Sekiya, M., \& Onishi, I.~K. 2018, The Astrophysical Journal, 860, \dodoi{10.3847/1538-4357/aac4a7}

\bibitem[{Senft \& Stewart(2008)}]{SenftStewart2008}
Senft, L.~E., \& Stewart, S.~T. 2008, Meteoritics and Planetary Science, 43, 1993

\bibitem[{Shibaike {et~al.}(2017)Shibaike, Okuzumi, Sasaki, \& Ida}]{Shibaikeetal2017}
Shibaike, Y., Okuzumi, S., Sasaki, T., \& Ida, S. 2017, The Astrophysical Journal, 846, \dodoi{10.3847/1538-4357/aa8454}

\bibitem[{Simon {et~al.}(2016)Simon, Armitage, Li, \& Youdin}]{Simonetal2016}
Simon, J.~B., Armitage, P.~J., Li, R., \& Youdin, A.~N. 2016, The Astrophysical Journal, 822, 55, \dodoi{10.3847/0004-637x/822/1/55}

\bibitem[{Simon {et~al.}(2022)Simon, Blum, Birnstiel, \& Nesvorn{\'y}}]{Simonetal2022}
Simon, J.~B., Blum, J., Birnstiel, T., \& Nesvorn{\'y}, D. 2022, Comets III, in press, \dodoi{10.48550/arXiv.2212.04509}

\bibitem[{Slattery {et~al.}(1992)Slattery, Benz, \& Cameron}]{Slatteryetal1992}
Slattery, W.~L., Benz, W., \& Cameron, A. G.~W. 1992, Icarus, 99, 167, \dodoi{10.1016/0019-1035(92)90180-F}

\bibitem[{Sossi {et~al.}(2018)Sossi, Moynier, \& van Zuilen}]{Sossietal2018}
Sossi, P.~A., Moynier, F., \& van Zuilen, K. 2018, Proc Natl Acad Sci U S A, 115, 10920, \dodoi{10.1073/pnas.1809060115}

\bibitem[{Stone \& Gardiner(2010)}]{StoneGardiner2010}
Stone, J.~M., \& Gardiner, T.~A. 2010, The Astrophysical Journal Supplement Series, 189, 142, \dodoi{10.1088/0067-0049/189/1/142}

\bibitem[{Stone {et~al.}(2008)Stone, Gardiner, Teuben, Hawley, \& Simon}]{Stoneetal2008}
Stone, J.~M., Gardiner, T.~A., Teuben, P., Hawley, J.~F., \& Simon, J.~B. 2008, The Astronomical Journal Supplement Series, 178, 137, \dodoi{10.1086/588755}

\bibitem[{Takahashi \& Inutsuka(2014)}]{Takahashietal2014}
Takahashi, S.~Z., \& Inutsuka, S.-i. 2014, The Astrophysical Journal, 794, \dodoi{10.1088/0004-637x/794/1/55}

\bibitem[{Takeuchi \& Lin(2002)}]{TakeuchiLin2002}
Takeuchi, T., \& Lin, D. N.~C. 2002, The Astronomical Journal, 581, 1344, \dodoi{10.1086/344437}

\bibitem[{Thiemens {et~al.}(2019)Thiemens, Sprung, Fonseca, Leitzke, \& Münker}]{Thiemensetal2019}
Thiemens, M.~M., Sprung, P., Fonseca, R. O.~C., Leitzke, F.~P., \& Münker, C. 2019, Nature Geoscience, 12, 696, \dodoi{10.1038/s41561-019-0398-3}

\bibitem[{Thiemens {et~al.}(2021)Thiemens, Tusch, O.~C.~Fonseca, Leitzke, Fischer-Gödde, Debaille, Sprung, \& Münker}]{Thiemensetal2021}
Thiemens, M.~M., Tusch, J., O.~C.~Fonseca, R., {et~al.} 2021, Nature Geoscience, 14, 716, \dodoi{10.1038/s41561-021-00821-1}

\bibitem[{Thompson \& Stevenson(1988)}]{ThompsonStevenson1988}
Thompson, C., \& Stevenson, D.~J. 1988, Astrophys. J., 333, 452, \dodoi{10.1086/166760}

\bibitem[{Thompson {et~al.}(2019)Thompson, Weinberger, Keller, Arnold, \& Stark}]{Thompsonetal2019}
Thompson, M.~A., Weinberger, A.~J., Keller, L.~D., Arnold, J.~A., \& Stark, C.~C. 2019, The Astrophysical Journal, 875, 45, \dodoi{10.3847/1538-4357/ab0d7f}

\bibitem[{Thompson \& Lauson(1972)}]{ThompsonLauson1972}
Thompson, S.~L., \& Lauson, H.~S. 1972, Sandia National Laboratories, Albuquerque, New Mexico, 119p

\bibitem[{Tominaga {et~al.}(2018)Tominaga, Inutsuka, \& Takahashi}]{Tominagaetal2018}
Tominaga, R.~T., Inutsuka, S.-i., \& Takahashi, S.~Z. 2018, Publications of the Astronomical Society of Japan, 70, 3, \dodoi{10.1093/pasj/psx143}

\bibitem[{Tominaga {et~al.}(2019)Tominaga, Takahashi, \& Inutsuka}]{Tominagaetal2019}
Tominaga, R.~T., Takahashi, S.~Z., \& Inutsuka, S.-i. 2019, The Astrophysical Journal, 881, 53, \dodoi{10.3847/1538-4357/ab25ea}

\bibitem[{Toro(1999)}]{Toro1999}
Toro, E.~F. 1999

\bibitem[{Touboul {et~al.}(2015)Touboul, Puchtel, \& Walker}]{Toubouletal2015}
Touboul, M., Puchtel, I.~S., \& Walker, R.~J. 2015, Nature, 520, 530, \dodoi{10.1038/nature14355}

\bibitem[{Wang \& Jacobsen(2016)}]{WangJacobsen2016}
Wang, K., \& Jacobsen, S.~B. 2016, Nature, \dodoi{10.1038/nature19341}

\bibitem[{Ward {et~al.}(2020)Ward, Canup, \& Rufu}]{Wardetal2020}
Ward, W.~R., Canup, R.~M., \& Rufu, R. 2020, Journal of Geophysical Research, 125, e2019JE00626, \dodoi{10.1029/2019JE006266}

\bibitem[{Weidenschilling(1977)}]{Weidenschilling1977}
Weidenschilling, S.~J. 1977, Monthly Notices of the Royal Astronomical Society, 180, 57, \dodoi{10.1093/mnras/180.2.57}

\bibitem[{Wiechert {et~al.}(2001)Wiechert, Halliday, Lee, Snyder, Taylor, \& Rumble}]{Wiecheretal2001}
Wiechert, U., Halliday, A.~N., Lee, D.~C., {et~al.} 2001, Science, 294, 345, \dodoi{10.1126/science.1063037}

\bibitem[{Williams \& Mukhopadhyay(2019)}]{WilliamsMukhopadhyay2019}
Williams, C.~D., \& Mukhopadhyay, S. 2019, Nature, 565, 78, \dodoi{10.1038/s41586-018-0771-1}

\bibitem[{Williams {et~al.}(2019)Williams, Mukhopadhyay, Rudolph, \& Romanowicz}]{Williamsetal2019}
Williams, C.~D., Mukhopadhyay, S., Rudolph, M.~L., \& Romanowicz, B. 2019, Geochemistry, Geophysics, Geosystems, 20, 4130, \dodoi{10.1029/2019gc008437}

\bibitem[{Yang {et~al.}(2017)Yang, Johansen, \& Carrera}]{Yangetal2017}
Yang, C.~C., Johansen, A., \& Carrera, D. 2017, Astronomy \& Astrophysics, 606, 80, \dodoi{10.1051/0004-6361/201630106}

\bibitem[{Yang \& Zhu(2021)}]{Yangetal2021}
Yang, C.-C., \& Zhu, Z. 2021, Monthly Notices of the Royal Astronomical Society, 508, 5538, \dodoi{10.1093/mnras/stab2959}

\bibitem[{Yokochi \& Marty(2004)}]{YokochiMarty2004}
Yokochi, R., \& Marty, B. 2004, Earth and Planetary Science Letters, 225, 77, \dodoi{10.1016/j.epsl.2004.06.010}

\bibitem[{Youdin(2011)}]{Youdin2011}
Youdin, A.~N. 2011, The Astrophysical Journal, 731, 99, \dodoi{10.1088/0004-637x/731/2/99}

\bibitem[{Youdin \& Goodman(2005)}]{YoudinGoodman2005}
Youdin, A.~N., \& Goodman, J. 2005, The Astrophysical Journal, 620, 459, \dodoi{10.1086/426895}

\bibitem[{Young {et~al.}(2016)Young, Kohl, Warren, Rubie, Jacobson, \& Morbidelli}]{Youngetal2016}
Young, E.~D., Kohl, I.~E., Warren, P.~H., {et~al.} 2016, Science, 351, 493, \dodoi{10.1126/science.aad0525}

\bibitem[{Zhang {et~al.}(2012)Zhang, Dauphas, M., Leya, \& Fedkin}]{Zhangetal2012}
Zhang, J., Dauphas, N., M., D.~A., Leya, I., \& Fedkin, A. 2012, Nature Geosci., 1429, 1, \dodoi{10.1038/NGEO1429}

\bibitem[{Ćuk {et~al.}(2021)Ćuk, Lock, Stewart, \& Hamilton}]{Cuketal2021}
Ćuk, M., Lock, S.~J., Stewart, S.~T., \& Hamilton, D.~P. 2021, The Planetary Science Journal, 2, 147, \dodoi{10.3847/PSJ/ac12d1}

\end{thebibliography}
\bibliographystyle{aasjournal}

\section*{Appendix}
\renewcommand{\theequation}{A.\arabic{equation}}
\setcounter{equation}{0}
The momentum equation for gas in the radial direction is written as
\begin{equation}
    \frac{v^2_{\phi}}{r}  = \frac{GM_\oplus}{r^2} + \frac{1}{\rho_{\rm g}}\frac{dp_{\rm g}}{dr},
\end{equation}
where $v_{\phi}$ is the azimuthal velocity of the gas. Using equation (\ref{eq:eta}) in the main text and the relationship of $p_{\rm g} = \rho_{\rm g} c_2^2$, 
\begin{equation}
v_{\phi} = v_{\rm K} (1-2 \eta)^{\frac{1}{2}}.
\label{eq:v_phi}
\end{equation}
The equations of motion of the particles in the gas in the radial and azimuthal directions, $v_{r}$ and $v_{\phi}$, are \citep{Armitage2010, TakeuchiLin2002}, 
\begin{equation}
\frac{dv_{ r}}{dt} = \frac{v^2_\phi}{r} - \Omega^2 r - \frac{1}{t_{\rm f}}(v_{ r}-v_{r, {\rm g}}),
\end{equation}
\begin{equation}
\frac{d }{dt}(rv_\phi) = - \frac{r}{t_{\rm f}}(v_\phi-v_{\phi,{\rm g}}),
\end{equation}
where $v_{r, {\rm g}}$ and $v_{ {\phi, {\rm g}}}$ are the gas velocities in the radial and azimuthal directions. Assuming that $\frac{d}{dt}$ terms are negligible, one finds that 
\begin{equation}
v_r = \frac{\tau^{-1}_{\rm f}v_{r, {\rm g}}-2\eta v_{\rm K}}{\tau_{\rm f} +\tau_{\rm f}^{-1} }.
\end{equation}
This formulation is slightly different from previous formulations ($-2\eta v_{\rm K}$ instead of $-\eta v_{\rm K}$) \citep{Armitage2010, TakeuchiLin2002} due to the different definition of $\eta$ by a factor of two. This leads to 
\begin{equation}
v_r = \frac{\tau^{-1}_{\rm f}v_{r, {\rm g}}-2\eta v_{\rm K}}{\tau_{\rm f} +\tau_{\rm f}^{-1} }.
\label{eq:vr}
\end{equation}
This can be further simplified by the values of $\tau_{\rm f}$ as
\begin{equation} 
\label{eq1}
\begin{split}
v_r
 & =     \begin{cases}
    v_{r,{\rm g}} - 2 \eta \tau_{\rm f} v_{\rm K} 
     & \text{at}\ \tau_{\rm f} \ll 1,  \\
          \frac{1}{2}(v_{r, {\rm g}}-2\eta v_{\rm K}) & \text{at}\  \tau_{\rm f} \sim 1, \\
     -\frac{2 \eta v_K}{\tau_{\rm f}}
       & \text{at}\  \tau_{\rm f} \gg 1. \\
    \end{cases}
\end{split}
\end{equation}
For a steady disk flow, one could assume $v_{r, {\rm g}} \sim  -\frac{3 \nu}{2r}$ \citep{Armitage2010}. In the main text, we simply assume $v_{r, {\rm g}}=0$.



\end{document}